\pdfoutput=1
\documentclass[pra,twocolumn,superscriptaddress,longbibliography]{revtex4-2}
\usepackage{graphicx,amsmath,amssymb}
\newenvironment{proof}[1][Proof]{\noindent\textbf{#1.} }{\ \rule{0.5em}{0.5em}}
\usepackage[colorlinks=true, citecolor=blue, urlcolor=blue]{hyperref}
\begin{document}
\title{Non-Markovian quantum thermometry}
\author{Ning Zhang}
\affiliation{Lanzhou Center for Theoretical Physics and Key Laboratory of Theoretical Physics of Gansu Province, Lanzhou University, Lanzhou 730000, China}
\author{Chong Chen}\email{chongchenn@gmail.com}
\affiliation{Department of Physics and The Hong Kong Institute of Quantum Information of Science and Technology, The Chinese University of Hong Kong, Shatin, New Territories, Hong Kong, China}
\author{Si-Yuan Bai}
\affiliation{Lanzhou Center for Theoretical Physics and Key Laboratory of Theoretical Physics of Gansu Province, Lanzhou University, Lanzhou 730000, China}
\author{Wei Wu}
\affiliation{Lanzhou Center for Theoretical Physics and Key Laboratory of Theoretical Physics of Gansu Province, Lanzhou University, Lanzhou 730000, China}
\author{Jun-Hong An}\email{anjhong@lzu.edu.cn}
\affiliation{Lanzhou Center for Theoretical Physics and Key Laboratory of Theoretical Physics of Gansu Province, Lanzhou University, Lanzhou 730000, China}
\begin{abstract}
The rapidly developing quantum technologies and thermodynamics have put forward a requirement to precisely control and measure the temperature of microscopic matter at the quantum level.	Many quantum thermometry schemes have been proposed. However, precisely measuring low temperature is still challenging because the obtained sensing errors generally tend to diverge with decreasing temperature. Using a continuous-variable system as a thermometer, we propose non-Markovian quantum thermometry to measure the temperature of a quantum reservoir. A mechanism to make the sensing error $\delta T$ scale with the temperature $T$ as the Landau bound $\delta T\simeq T$ in the full-temperature regime is discovered. Our analysis reveals that it is the quantum criticality of the total thermometer-reservoir system that causes this enhanced sensitivity. Efficiently avoiding the error-divergence problem, our result gives an efficient way to precisely measure the low temperature of quantum systems.
\end{abstract}
\maketitle
%%%%%%%%%%%%%%%%%%%%%%%%%%%%%%%%%%%%%%%%%%%%%%%%%%%%%%%%%%%%%%%%%%
\section{Introduction}
Precise sensing of temperature is of significance in fields ranging from the fundamental natural sciences to the rapidly developing quantum technologies \cite{RevModPhys.78.217,10.1039/9781782622031, DePasquale2018, Mehboudi2019}. People's increased capabilities of controlling and using quantum characters of microscopic matter have led to the development of the field of quantum thermodynamics \cite{Campisi2011,Brandao2015,doi:10.1080/00107514.2016.1201896,10.1088/2053-2571/ab21c6}, where the precise measuring of thermodynamic quantities at the quantum level invalidates classical sensing schemes and calls for advanced ones. On the other hand, temperature is one of the main reasons causing decoherence, which is a main bottleneck in the practical realization of quantum technology protocols. Thus, from an application viewpoint, quantum devices generally work at ultralow temperature, e.g., cold-atom and ion-trap systems \cite{PhysRevA.88.063609,OLF2015,Mehboudi2019a,PhysRevX.10.011018,PhysRevLett.125.080402}, which also calls for the ability to precisely control and sense temperature.

Quantum thermometry aims to realize the precise measurement of temperature using quantum features \cite{PhysRevA.82.011611,Brunelli2011,Jevtic2015,Correa2015, Hofer2017,Correa2017,Campbell_2017,Kiilerich2018,Feyles2019,Mukherjee2019,Potts2019fundamentallimits,PhysRevResearch.2.033394,Montenegro2020,Gebbia2020,planella2021bathinduced,PhysRevResearch.1.033021, DePasquale2017, Cavina2018, Seah2019, PhysRevLett.125.080402}. A quantum system is chosen as a thermometer and is brought into thermal contact with the measured system in thermal equilibrium. The temperature is measured in either the equilibrium thermal state or nonequilibrium dynamical state of the thermometer through certain observables. It has been explored in various platforms, including nitrogen-vacancy centers in diamond \cite{Toyli8417,Kucsko2013}, optical nanofibres \cite{PhysRevA.92.013850}, cavity optomechanical systems \cite{purdy2017quantum}, and quantum dots \cite{PhysRevApplied.2.024001}. The advantage of using quantum features in thermometry is that an enhanced precision can be achieved in certain temperature regimes due to quantum coherence \cite{PhysRevA.82.011611,Brunelli2011,Jevtic2015}, strong coupling \cite{Correa2017,Mehboudi2019a}, quantum correlation \cite{purdy2017quantum,Gebbia2020,planella2021bathinduced}, periodic driving \cite{Mukherjee2019}, or nonequilibrium dynamics \cite{PhysRevResearch.1.033021, DePasquale2017, Cavina2018, Seah2019, PhysRevLett.125.080402}. A challenge of almost all of the existing schemes is that their sensing errors tend to diverge with decreasing temperature \cite{Pasquale2016,PhysRevA.95.052115}, although some ways to slow down the divergence via strong coupling \cite{Mehboudi2019a}, periodic driving \cite{Mukherjee2019}, and finite measurement resolution \cite{Potts2019fundamentallimits,PhysRevResearch.2.033394} have been explored. Therefore, quantum thermometry performing well in the low-temperature regime is still absent.

In this paper, we propose a quantum thermometry scheme efficiently avoiding the error-divergence problem in the low-temperature regime. Using a continuous-variable system as a thermometer to measure the temperature of a quantum reservoir, our scheme permits us to achieve a scaling of sensing error as $\delta T\simeq T$, which is called the Landau bound \cite{Paris_2015}, in the full-temperature regime. We find that it is the combined action of the non-Markovian dynamical encoding of the thermometer and quantum criticality of the total thermometer-reservoir system that causes this notable performance. The quantum criticality occurs as a consequence of quantum phase transition of the total thermometer-reservoir system with the abrupt formation of a bound state out of its continuous energy band. Supplying a useful idea to design quantum thermometry, our study may potentially prompt advances of low-temperature sensing in quantum thermodynamics and technologies.

\section{Quantum thermometry} \label{TE}
Quantum sensing to a quantity $\theta$ of a certain system generally involves three steps \cite{RevModPhys.89.035002,RevModPhys.90.035005}. One first prepares a quantum sensor in certain state $\rho_\text{in}$. Then the interaction of the sensor with the measured system is switched on to encode $\theta$ into the sensor state $\rho_\theta=\check{\mathcal E}\rho_\text{in}$. Acting on the Liouvillian space of the density matrix, the superoperator $\check{\mathcal E}$ may be either unitary or nonunitary. Lastly, one measures the sensor and infers the value of $\theta$ from the results. The inevitable errors mean that one cannot estimate $\theta$ exactly. The ultimate estimation error of $\theta$ is constrained by the quantum Cram\'{e}r-Rao bound $\delta \theta=(\mathcal{N}\mathcal{F}_\theta)^{-1/2}$  \cite{RevModPhys.90.035006,Liu_2019}. Here $\mathcal{N}$ is the measurement times and $\mathcal{F}_\theta=\text{Tr}(\hat{L}_\theta^2\rho_\theta)$, with $\hat{L}_\theta$ determined by $\partial_\theta\rho_\theta=(\hat{L}_\theta \rho_\theta+\rho_\theta\hat{L}_\theta)/2$, is the quantum Fisher information (QFI) describing the most information for estimating $\theta$ from $\rho_\theta$. Due to the independence of $\mathcal{F}_\theta$ on $\mathcal{N}$, we choose $\mathcal{N}=1$. It has been found that entanglement \cite{Wang_2017,Hosten}, squeezing \cite{PhysRevLett.123.231107, PhysRevLett.123.231108,PhysRevApplied.13.024037,PhysRevLett.123.040402}, chaos \cite{Fiderer2018}, and criticality \cite{PhysRevA.78.042105,PhysRevLett.121.020402,Salado_Mej_a_2021} can act as quantum resources to beat the sensitivity limit of classical sensors.

We are interested in measuring the temperature of a quantum reservoir with infinite degrees of freedom. We choose a continuous-variable system as the quantum thermometer. It may be an $LC$ oscillator \cite{Hofer2017}, harmonic potential trapped BEC \cite{Mehboudi2019}, or mechanical oscillator \cite{Aspelmeyer2014}. Its interaction with the reservoir for encoding the temperature into the state of the thermometer reads ($\hbar=1$)
\begin{equation}
	\hat{H}=\omega_{0}\hat{a}^{\dagger} \hat{a}+\sum_{k} [\omega_{k}\hat{b}^{\dagger}_{k} \hat{b}_{k} +g_{k} (\hat{a}^{\dagger} \hat{b}_{k}+\hat{b}^{\dagger}_{k}\hat{a})],\label{Hamlt}
\end{equation}
where $\hat{a}$ and $\hat{b}_{k}$ are the annihilation operators of the thermometer with frequency $\omega_{0}$ and the $k$th reservoir mode with frequency $\omega_{k}$, and $g_k$ is their coupling strength. Their coupling is further characterized by the spectral density $J(\omega)=\sum_{k} g^{2}_{k} \delta (\omega-\omega_{k})$. We consider the Ohmic-family spectral density $J(\omega)=\eta\omega^{s}\omega_{c}^{1-s} e^{-\omega/\omega_{c}}$, where $\eta$ is a dimensionless coupling constant, $\omega_{c}$ is a cutoff frequency, and $s$ is an Ohmicity index \cite{RevModPhys.59.1}. It is widely used to describe the noises in circuit QED \cite{PhysRevLett.97.016802,Forn-Diaz2017,RevModPhys.86.361}, ion trap  \cite{PhysRevA.78.010101}, and waveguide \cite{PhysRevLett.120.153602} systems. Depending on the dispersion relation and density of states of the reservoir, $\omega_c$ characterizes the dominate modes coupled to the thermometer and relates to the typical time scale of the correlation function of the reservoir. The index $s$ generally is determined by the reservoir dimension \cite{Weiss}. The temperature $T$ is carried by the initial state $\rho_\text{R}(0)=\prod_{ k}e^{-\beta\omega_{k} \hat{b}^{\dagger}_{k} \hat{b}_{k}}/\text{Tr}[e^{-\beta\omega_{k} \hat{b}^{\dagger}_{k} \hat{b}_{k}}]$, where $\beta=(K_{B}T)^{-1}$ and $K_{B}$ is the Boltzmann constant.

Setting the initial state of the total system as $\rho _\text{tot}(0)=\rho(0)\otimes \rho_\text{R}(0)$, we can derive the exact non-Markovian master equation of the quantum thermometer by the path-integral influence-functional method \cite{Feynman1963,An2007,Yang2014} as
\begin{eqnarray}
\dot{\rho}(t) &=&-i\Omega (t)[\hat{a}^{\dag }\hat{a},\rho (t)] +[\Gamma (t)+{\Gamma ^{\beta }(t)}/{2}]\check{\mathcal{L}}_{\hat{a}}\rho(t)  \nonumber \\
&&+{\Gamma ^{\beta }(t)}/{2}\check{\mathcal{L}}_{\hat{a}^\dag}\rho(t), \label{ExM}
\end{eqnarray}where $\check{\mathcal{L}}_{\hat{o}}\cdot=2\hat{o}\cdot\hat{o}^\dag-\cdot\hat{o}^\dag\hat{o}-\hat{o}^\dag\hat{o}\cdot$ is the Lindblad superoperator.
Here $\Omega (t)=-\textrm{Im}[\dot{u}(t)/u(t)]$ is the renormalized frequency, $\Gamma (t)=-\textrm{Re}[\dot{u}(t)/u(t)]$ and $\Gamma ^{\beta }(t)=\dot{v}(t)+2v(t)\Gamma(t)$ are the dissipation and noise coefficients. The functions $u(t)$ and $v(t)$ are determined by
\begin{eqnarray}
&\dot{u}(t)+i\omega _{0}u(t)+\int_{0}^{t}dt_{1}\mu(t-t_{1})u(t_{1}) =0, & \label{u}\\
&v(t)=\int_0^t dt_1\int_0^tdt_2 u^*(t_1)\nu(t_1-t_2)u(t_2),\label{extv}&
\end{eqnarray}
under $u(0)=1$. The kernel functions $\mu(x)=\int_{0}^{\infty }d\omega J(\omega )e^{-i\omega x}$ and $\nu(x)=\int_{0}^{\infty }d\omega J(\omega )\bar{n}(\omega)e^{-i\omega x}$ with $\bar{n}(\omega)=(e^{\beta \omega }-1)^{-1}$.
Keeping the same Lindblad form as the Born-Markovian master equation \cite{breuer2002theory}, Eq. (\ref{ExM}) incorporates all the non-Markovian effects induced by the backactions of the reservoir into these time-dependent coefficients self-consistently. It can be seen that the temperature of the reservoir is successfully encoded into the thermometer state $\rho(t)$ via $\Gamma^\beta(t)$. It is worth noting that the encoding in our scheme is different from that of either quantum metrology schemes based on Ramsey \cite{PhysRevLett.79.3865} and Mech-Zehnder \cite{PhysRevD.23.1693} interferometers, where the encoding dynamics is unitary, or quantum sensing to the spectral density of a reservoir \cite{Wu2021,PhysRevApplied.15.054042}, where, although the encoding is also nonunitary, the sensed quantities are carried by the sensor-reservoir interactions. %The unique character of quantum thermometry resides in that the thermal fluctuation governed by the finite temperature makes the inherently fragile quantum features hard to play a same constructive role as other quantum sensors \cite{RevModPhys.89.035002,RevModPhys.90.035005}

We consider that the initial state of the thermometer is a coherent state, i.e., $\rho(0)=|\alpha_0\rangle\langle \alpha_0|$. Governed by Eq. \eqref{ExM}, it evolves to (see Appendix \ref{drstd})
\begin{equation}
\rho(t)=\sum_{n=0}^\infty{M(t)^{n+1}v(t)^n}\hat{\mathcal{D}}_t|n\rangle\langle n|\hat{\mathcal{D}}_t^\dag,\label{tmdst}
\end{equation}
where $M(t)=[1+v(t)]^{-1}$ and $\hat{\mathcal{D}}_t=\exp[\alpha_0 u(t)\hat{a}^\dag-\alpha_0^*u^*(t)\hat{a}]$. As a Gaussian state, the characteristic function of Eq. \eqref{tmdst} is of Gaussian form \cite{RevModPhys.77.513}:
$\chi ({\pmb\gamma })\equiv\text{Tr}[\rho \hat{\mathcal D}({\gamma})]=\exp(-{\frac{1}{4}}{\pmb\gamma }^{\dag }{\pmb \sigma} {\pmb\gamma}-i{\mathbf{d}}^{\dag }K{\pmb\gamma })$, where ${\pmb \gamma}=(\gamma,\gamma^*)^T$, $K=\text{diag}(1,-1)$, and the elements of the displacement vector ${\mathbf{d}}$ and the covariant
matrix ${\pmb \sigma}$ are $d_{i} =\text{Tr}(\rho \hat{A}_{i})$ and $\sigma _{ij} =\text{Tr}[\rho \{\Delta \hat{A}_{i},\Delta \hat{A}_{j}^{\dag}\}]$
with $\hat{\mathbf{A}}=(\hat{a},\hat{a}^\dag)^T$ and $\Delta \hat{A}_{i}=\hat{A}_{i}-d_{i}$. Its QFI for $\theta$ reads
$\mathcal{F}_{\theta}=\frac{1}{2}[\text{vec}(\partial_\theta {\pmb \sigma})]^\dag\mathcal{M}^{-1}\text{vec}(\partial_\theta {\pmb \sigma})+2(\partial_{\theta}\mathbf{d})^{\dagger}\pmb{\sigma}^{-1}\partial_{\theta}\mathbf{d}$, where $\mathcal{M}={\pmb\sigma}^*\otimes{\pmb\sigma}-K\otimes K$, with $\pmb{\sigma}^{*}$ being the complex conjugate of $\pmb{\sigma}$ \cite{_afr_nek_2015}. The QFI of the temperature for Eq. \eqref{tmdst} reads
\begin{equation}
\mathcal{F}_{T}(t)={M(t)[\partial_T v(t)]^2/v(t)}.\label{stdst}
\end{equation}
It can be proven that the QFI \eqref{stdst} is reached by measuring the number operator $\hat{a}^\dag\hat{a}$ of the thermometer.

As a special case, we first revisit the precision under the Born-Markovian approximation. When the coupling is weak and the time scale of the reservoir correlation function is much smaller than that of the thermometer, we can make this approximation and obtain $u_\text{MA}(t)=e^{-[\kappa+i(\omega_0+\Delta(\omega_0))]t}$ and $v_\text{MA}(t)=\bar{n}(\omega_0)(1-e^{-2\kappa t})$ (Appendix \ref{DrBMas}). Here $\kappa=\pi J(\omega_0)$ and $\Delta(\omega_0)=\mathcal{P}\int{J(\omega)\over \omega_0-\omega}d\omega$, with $\mathcal{P}$ denoting the Cauchy principal value. Then the coefficients in Eq. (\ref{ExM}) reduce to $\Gamma_\text{MA}(t)=\kappa$, $\Omega_\text{MA}(t)=\omega_0+\Delta(\omega_0)$, and $\Gamma_\text{MA}^\beta(t)=2\kappa \bar{n}(\omega_0)$. The unique steady state is a canonical state $\rho_\text{MA}(\infty)=e^{-\beta\omega_0\hat{a}^\dag\hat{a}}/\text{Tr}[e^{-\beta\omega_0\hat{a}^\dag\hat{a}}]$, which is independent of the initial state. Thus the thermometer equilibrates to a thermal state with the same temperature as the reservoir in Born-Markovian dynamics. According to Eq. \eqref{stdst}, we obtain the QFI
\begin{equation}
\mathcal{F}_{T}^\text{MA}(t)=\bar{ F}_T(\omega_0){\bar{n}(\omega_0)+1\over \bar{n}(\omega_0)+(1-e^{-2\kappa t})^{-1} },\label{MAFI}
\end{equation}where $\bar{ F}_T(\omega_0)\equiv(\beta\omega_0)^2 \bar{n}(\omega_0)[1+\bar{n}(\omega_0)]/T^2$. We have neglected the constant $\Delta(\omega_0)$, which is generally renormalized into $\omega_0$ \cite{RevModPhys.59.1}. Equation \eqref{MAFI} reveals that the QFI increases with time and saturates to $\bar{F}_T$ corresponding to the QFI of $\rho_\text{MA}(\infty)$. The equilibrium-state performance of the thermometer reads $\lim_{T\rightarrow\infty}\bar{ F}_T(\omega_0)=T^{-2}$ in the high-temperature regime, which is called the Landau bound \cite{Paris_2015}. However, it is unfortunate to find that the QFI tends to zero in the low-temperature limit. Being consistent with previous works \cite{PhysRevA.82.011611,Brunelli2011,Jevtic2015,Correa2015, Hofer2017,Correa2017,Campbell_2017,Kiilerich2018,Feyles2019,Mukherjee2019,Potts2019fundamentallimits,PhysRevResearch.2.033394,Montenegro2020,Gebbia2020,planella2021bathinduced,PhysRevResearch.1.033021, DePasquale2017, Cavina2018, Seah2019, PhysRevLett.125.080402}, this means that the thermometer becomes insufficient for measuring low temperatures under the Born-Markovian approximation.

The non-Markovian solution of Eq. \eqref{u} is \cite{Wu2021}
\begin{equation}
u(t)=Ze^{-iE_{b}t}+\int_{0}^{\infty}dE\Theta(E)e^{-iE t},
\label{uspe}
\end{equation}
with $Z=[1+\int_{0}^{\infty}\frac{J(\omega)}{(E_{b}-\omega)^{2}}d\omega]^{-1}$ and $\Theta(E)=\frac{J(E)}{[E-\omega_{0}-\Delta(E)]^{2}+[\pi J(E)]^{2}}$. Here $E_b$ is a possibly formed isolated root of the transcendental equation
\begin{equation}~\label{transeq}
y(E)\equiv\omega_{0}-\int_{0}^{\infty}\frac{J(\omega)}{\omega-E}d\omega=E.
\end{equation}
The solutions of Eq. \eqref{transeq} are the eigenenergies of Eq. \eqref{Hamlt} in the single-excitation subspace. Since $y(E)$ is a decreasing function in the regime $E<0$, Eq.~(\ref{transeq}) has one isolated root $E\equiv E_{b}$ provided $y(0)<0$. We call the eigenstate corresponding to this isolated eigenenergy the bound state. On the contrary, it has infinite roots in the regime $E>0$, which form a continuous energy band. Since an extra band gap is induced by the formation of the bound state, we claim that a quantum phase transition occurs in the total system. Such a quantum phase transition has a profound impact on the dynamics of the thermometer. Contributed by the continuous energy band, the integral in Eq. \eqref{uspe} gradually vanishes with time due to out-of-phase interference. Thus, if the bound state is formed, then $\lim_{t\rightarrow\infty}u(t)=Ze^{-iE_{b}t}$, leading to a dissipationless dynamics; while if it is absent, then $\lim_{t\rightarrow\infty}u(t)=0$, characterizing a complete decoherence. It has been found that, when the bound state is absent, the thermometer equilibrates to a thermal state at an effective temperature; whenever the bound state is present, it tends to a steady state no longer capable of being described by thermal state \cite{Yang2014}.   For the Ohmic-family spectral density, the bound state is formed when $\omega_{0}/\omega_{c}<\eta\gamma(s)$, where $\gamma(s)$ is Euler's $\gamma$ function. Different from the previous schemes based on the thermal state \cite{PhysRevA.82.011611,Campbell_2017,Potts2019fundamentallimits,PhysRevResearch.1.033021,Pasquale2016,PhysRevB.98.045101,PhysRevA.95.052115,Paris_2015}, our dynamical scheme enables us to reveal its full performance in both the transient process and steady state regardless of whether the thermometer equilibrates to a thermal state or not.

\begin{figure}[tbp]
\begin{center}
  \includegraphics[width=0.90\columnwidth]{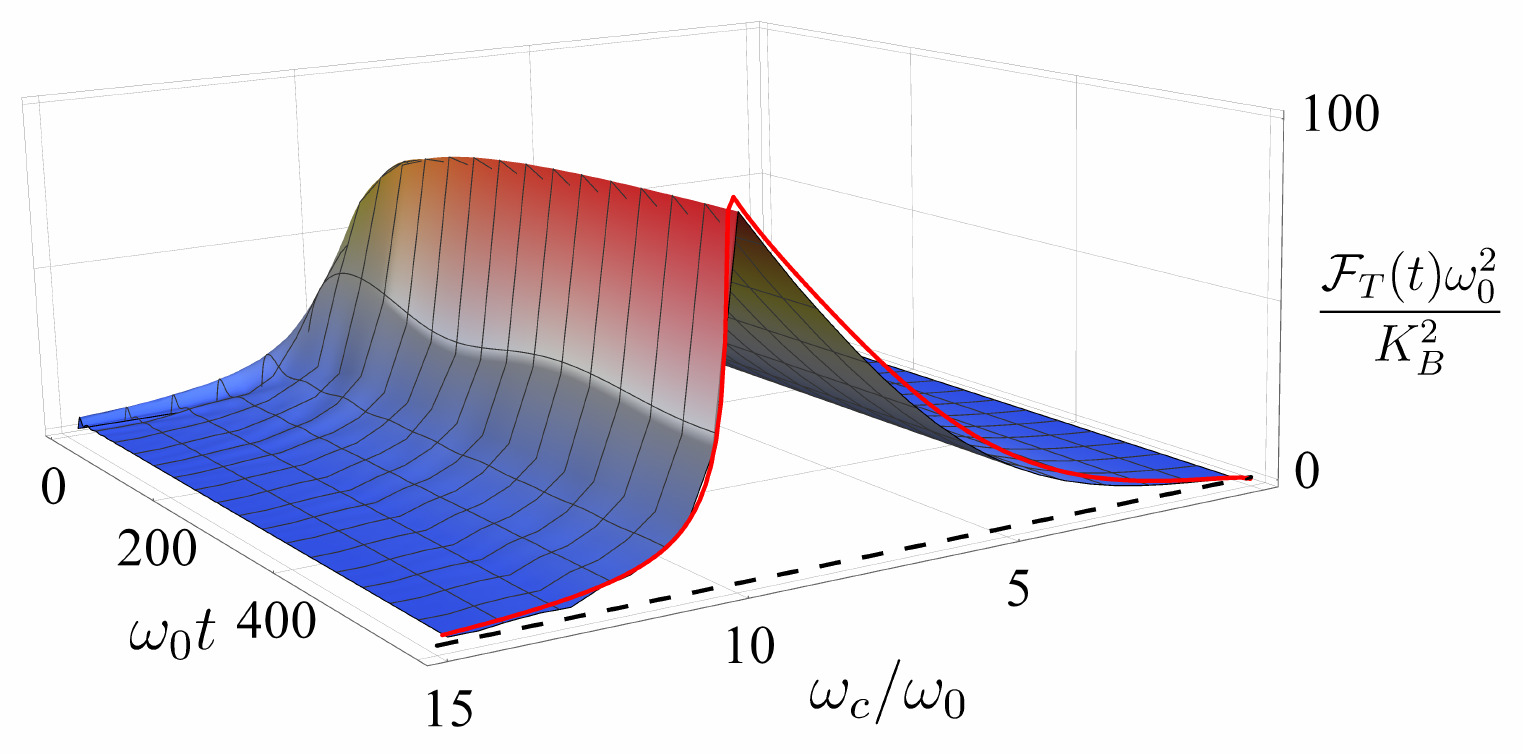}\\
  \caption{Evolution of the non-Markovian QFI for different $\omega_c$ by numerically solving Eqs. \eqref{u} and \eqref{extv}. The red solid line is analytically obtained from Eq. \eqref{stdqf} and the black dashed line is the long-time QFI $\bar{{F}}_T(\omega_0)$ under the Born-Markovian approximation. We use $\eta=0.1$, $s=1$, and $T=0.1\omega_0/K_B$.}
  \label{timedep}
\end{center}
\end{figure}

Equation \eqref{extv} is recast into $v(t)=\int^{\infty}_{0} d\omega  A_{\omega}(t) \bar{n}(\omega)$, where $A_\omega(t)=J(\omega)|\tilde{u}_\omega(t)|^2$, with $\tilde{u}_\omega(t)=\int_0^t d\tau u(\tau)e^{i\omega\tau}$, is called the heat-exchange spectrum, which tends to $A_\omega(\infty)=\Theta(\omega)+{Z^2J(\omega)/ (\omega-E_b)^2}$. Then we obtain from Eq. \eqref{stdst} an upper bound of the QFI (see Appendix \ref{drqfi})
\begin{equation}
\mathcal{F}_T(\infty)\leq M(\infty){\int_0^\infty d\omega \bar{{F}}_T(\omega)A_\omega(\infty)[1+\bar{n}(\omega)]}.\label{stdqf}
\end{equation}
This reveals that the non-Markovian QFI is fully determined by the overlap integral between the QFI of an equilibration mode with given frequency $\omega$ and the thermal-excitation dressed heat-exchange spectrum $M(\infty) A_{\omega}(\infty) [1+\bar{n}(\omega)]$. First,  $\lim_{T\rightarrow\infty}\bar{{F}}_T(\omega)=T^{-2}$ reduces Eq. \eqref{stdqf} to $\lim_{T\rightarrow\infty}\mathcal{F}_T(\infty)\leq [1-Z^2M(\infty)]T^{-2}$ in the high-temperature limit, where $\int_0^\infty d\omega A_\omega(\infty)=1-Z^2$ has been used. This scaling relation with temperature matches with the Landau bound \eqref{MAFI} under the Born-Markovian approximation except for the prefactor when the bound state is formed. Second, it is remarkable to find that $v(\infty)$ and $A_{\omega}(\infty)$ show an infrared divergence at the critical point of forming the bound state. Therefore, the integral in Eq. \eqref{stdqf} is dominated by the infrared-frequency regime. Using $\lim_{\omega\rightarrow 0}\bar{{F}}_T(\omega)=T^{-2}$ at the critical point, we readily convert Eq. \eqref{stdqf} into \begin{equation}\mathcal{F}_T(\infty)|_\text{CP}\leq T^{-2}\label{marst}\end{equation} at the critical point of forming the bound state (see Appendix \ref{drqfi}). This scaling relation works well in the full-temperature regime. Such quantum-criticality-enhanced QFI succeeds in avoiding the problem that the QFI tends to zero in the low-temperature regime in conventional quantum thermometry schemes \cite{PhysRevA.82.011611,Brunelli2011,Jevtic2015,Correa2015, Hofer2017,Correa2017, Campbell_2017, Kiilerich2018,Feyles2019,Mukherjee2019,Potts2019fundamentallimits,PhysRevResearch.2.033394,Montenegro2020,Gebbia2020,planella2021bathinduced,PhysRevResearch.1.033021, DePasquale2017, Cavina2018, Seah2019, PhysRevLett.125.080402}. Being independent of $s$, our low-temperature scaling relation surpasses the result $\mathcal{F}_T\propto T^{s-1}$ obtained in the spin-boson model \cite{PhysRevResearch.2.033394}. Note that our quantum-criticality-enhanced thermometry is substantially different from that in Ref. \cite{PhysRevB.98.045101}, where a similar scaling is obtained for the Caldeira-Leggett model without quantum criticality under the condition of $\omega_0$ equal to zero.

\begin{figure}[tbp]
\begin{center}
  \includegraphics[width=\columnwidth]{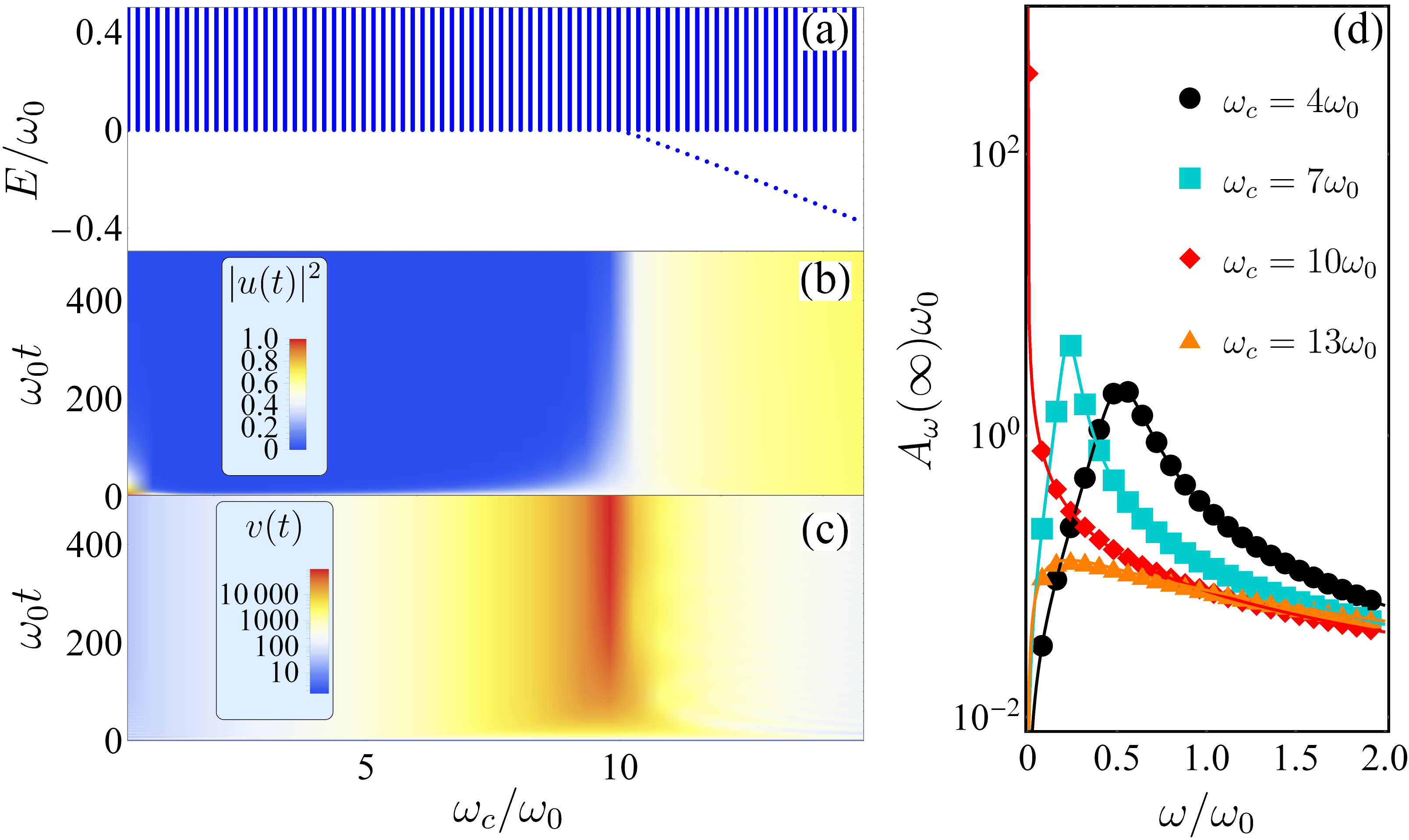}\\
  \includegraphics[width=0.90\columnwidth]{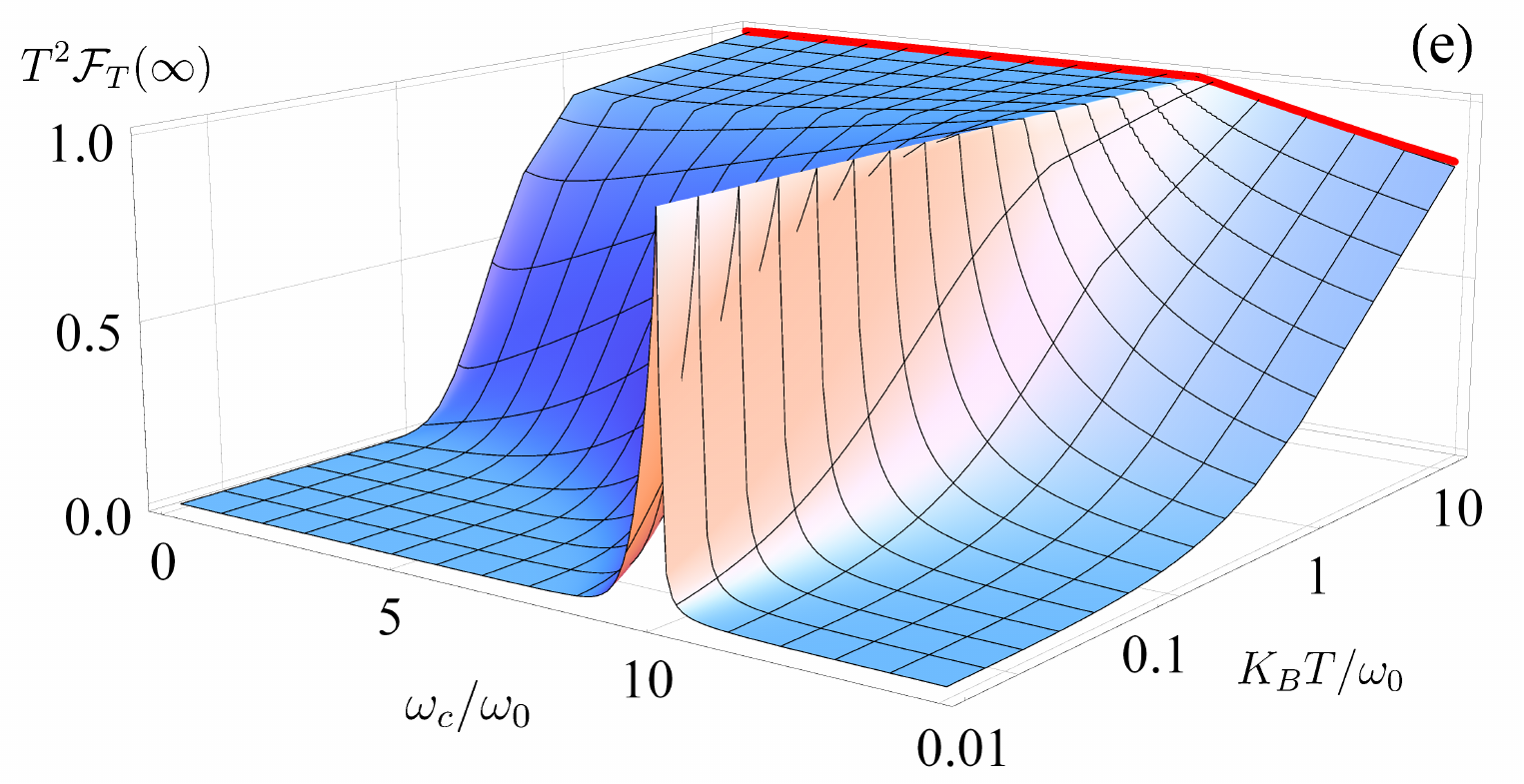}
  \caption{Energy spectrum in the single-excitation space of the total system (a), $|u(t)|^2$ (b), and $v(t)$ (c) as a function of $\omega_c$. (d) Heat-exchange spectrum $A_{\omega}(\infty)$ for different $\omega_{c}$, which shows an infrared divergence at the critical point $\omega_0=\eta\omega_c$. (e) Steady-state QFI for different $T$ and $\omega_c$. The red solid line is obtained via the analytical function $[1-Z^2M(\infty)]T^{-2}$. Other parameters are the same as Fig. \ref{timedep}. }
  \label{crtc}
\end{center}
\end{figure}

\section{Numerical calculations} \label{NS}
Taking the Ohmic spectral density as an example, we plot in Fig. \ref{timedep} the non-Markovian evolution of $\mathcal{F}_T(t)$ for different cutoff frequencies $\omega_c$. It can be seen that $\mathcal{F}_T(t)$ gradually increases with time from zero to $\omega_c$-dependent stable values, which are larger than the Markovian approximate one $\bar{F}_T(\omega_0)$ in the full-parameter regime. It indicates one of the advantages of our non-Markovian quantum thermometer over the Markovian approximate ones. The stable $\mathcal{F}_T(\infty)$ shows good matching with the analytical form in Eq. \eqref{stdqf}. It verifies the validity of Eq. \eqref{stdqf} in characterizing the steady-state performance of our quantum thermometer. Another interesting feature is that an obvious maximum of the QFI is present at $\omega_0=\eta\omega_c$. To uncover the physical reason, we plot in Fig. \ref{crtc}(a) the energy spectrum in the single-excitation space of the total system of Eq. \eqref{Hamlt}. We really see that an isolated eigenenergy with the associated eigenstate called the bound state is present in the band-gap regime when $\omega_0<\eta\omega_c$. Its presence opens an extra band gap in the energy spectrum, which signifies a quantum phase transition of the total system. Accompanying its formation, the long-time $|u(t)|^2$ abruptly increases from zero to a finite value [see Fig. \ref{crtc}(b)]. The functions $v(t)$ and $A_\omega(\infty)$, respectively, show a long-time and low-frequency divergence at the critical point of forming the bound state [see Figs. \ref{crtc}(c) and \ref{crtc}(d)]. Because $\bar{{F}}_{T}(\omega) (\bar{n}(\omega)+1)$ is a decreasing function of $\omega$, the overlap integral in Eq. \eqref{stdqf} is dominated by the low frequencies and leads to the maximum at the critical point. All the results confirm that the maximum of $\mathcal{F}_T(\infty)$ at $\omega_0=\eta\omega_c$ is intrinsically rooted in the quantum criticality induced by the bound state.

\begin{figure}[t]
\begin{center}
 \includegraphics[width=\columnwidth]{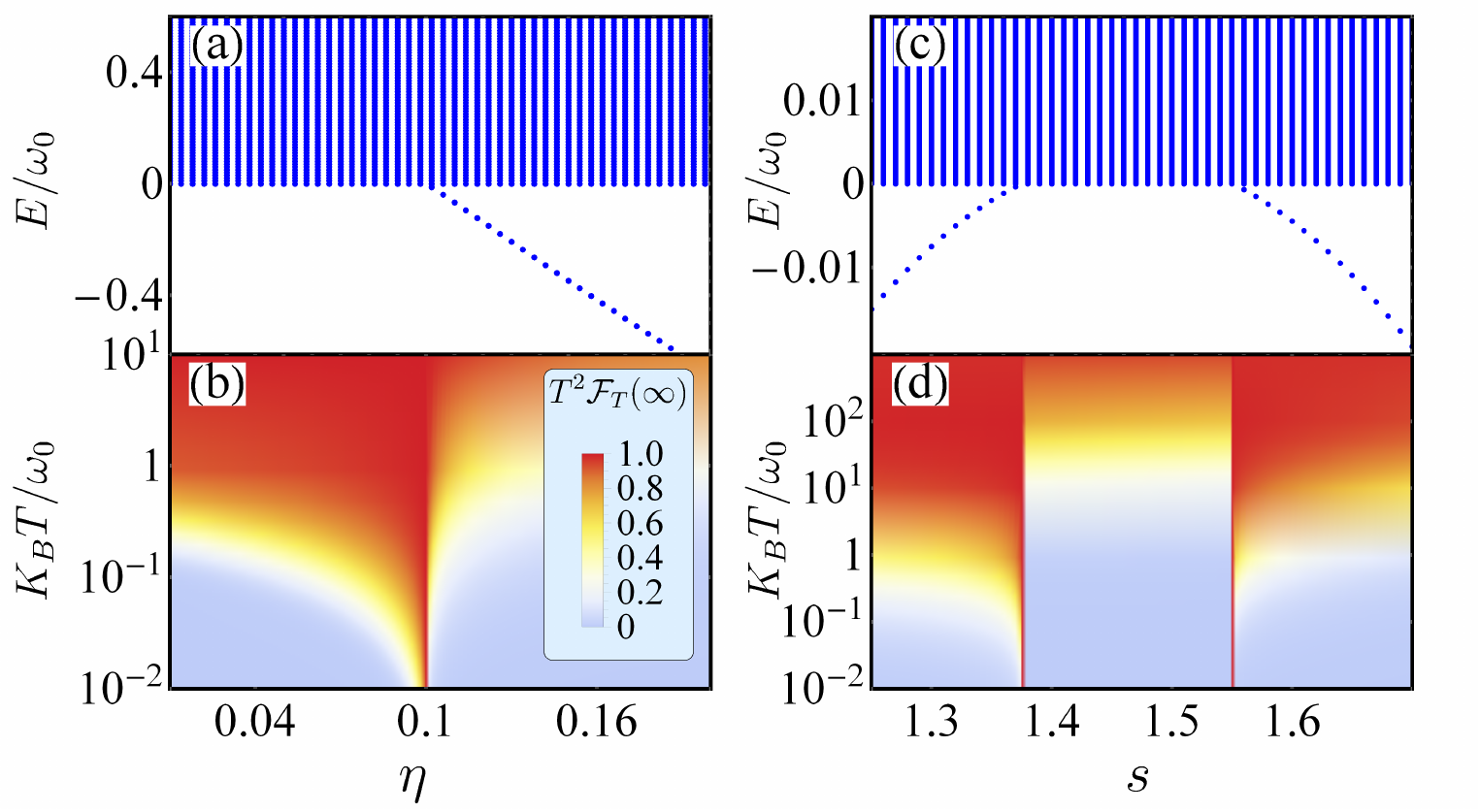}
  \caption{Energy spectra for different $\eta$ when $s=1$ (a) and for different $s$ when $\eta=0.1125$ (c). (b), (d) The corresponding steady-state QFI for different $T$. $\omega_c=10\omega_0$ is used.}
  \label{stdars}
\end{center}
\end{figure}
To further reveal the superiority of such quantum-criticality-enhanced thermometry, we plot in Fig. \ref{crtc}(e) the steady-state QFI $\mathcal{F}_T(\infty)$ for different $T$ and $\omega_c$. It clearly demonstrates that, in the high-temperature regime, $\mathcal{F}_T(\infty)$ matches with our analytical result $[1-Z^2M(\infty)]T^{-2}$. At the critical point of forming the bound state, the QFI scales with the temperature as Eq. \eqref{stdqf} in the full-temperature regime, which means that the performance of our non-Markovian quantum thermometry becomes better and better with a decrease of the temperature. A similar performance can be obtained by tuning the coupling constant $\eta$ [see Figs. \ref{stdars}(a) and \ref{stdars}(b)]. This successfully solves the problem of the conventional schemes where the QFI tends to zero in the low-temperature regime. It is noted that the sensitivity \eqref{stdqf} is achievable not only exactly at the critical point, but also in a relatively wide parameter regime near the critical point. Via the criticality-scaling analysis, we find that the condition to achieve \eqref{stdqf} can be relaxed to $|\omega_0-\eta\omega_c|\leq 1.52K_BT$ (see Appendix \ref{critpndt}). Thus, the parameter regime supporting \eqref{stdqf} becomes
wider with increasing temperature, as confirmed by Figs. \ref{crtc}(e) and \ref{stdars}(b). Given the fact that we never experiment exactly at zero temperature, it endows our scheme with a fault tolerance to the imprecise parameter tuning to reach the critical point. This parameter regime also reveals that what really matters is the relative value $\omega_c/\omega_0$. The critical regime is also achievable by tuning $\omega_0$ for given $\omega_c$ and $\eta$.

\section{Discussion and conclusions}
Our result can be generalized to other forms of spectral density, where the specific condition on the quantum criticality may be different, but the conclusion remains unchanged [see Figs. \ref{stdars}(c) and \ref{stdars}(d)]. Many ways of controlling the spectral density in circuit QED \cite{PhysRevLett.97.016802,Forn-Diaz2017,RevModPhys.86.361} and trapped ion \cite{PhysRevA.78.010101} platforms have been proposed. As the essential feature of our scheme, the bound state and its dynamical effect have been observed in circuit QED \cite{Liu2016} and ultracold atom \cite{Kri2018} systems. These progresses indicate that our scheme via engineering the quantum criticality caused by the bound state is realizable in the state-of-the-art technique of quantum optics experiments.

In summary, we propose a non-Markovian quantum thermometry scheme to measure the equilibrium temperature of a quantum reservoir by use of a continuous-variable system as a thermometer. A quantum criticality induced by the formation of a thermometer-reservoir bound state and its crucial role in enhancing the performance of the thermometer are revealed. It is revealed that the scaling relation of the QFI in the long-time limit reaches the Landau bound $T^{-2}$ in the full-temperature regime near the critical point. This efficiently avoids the problem that the QFI of conventional quantum thermometry schemes tends to zero in the low-temperature regime. Supplying a mechanism for designing high-precision sensors to low temperature, our result may be used in developing temperature-monitoring components in quantum thermodynamics and quantum device fabrication.

\section*{Acknowledgments}
The work is supported by the National Natural Science Foundation (Grants No. 11875150, No. 11834005, and No. 12047501).

\appendix
\section{Time-dependent state of the thermometer}\label{drstd}
Here, we give the derivation details of Eq. \eqref{tmdst}. According to the path-integral influence-function theory, the evolution of the thermometer state in the coherent-state representation \cite{Yang2014} reads
\begin{eqnarray}
\rho (\bar{\alpha}_{f},\alpha _{f}^{\prime };t) &=&\int d\mu (\alpha_{i})d\mu (\alpha _{i}^{\prime })\mathcal{J}(\bar{\alpha}_{f},\alpha_{f}^{\prime };t|\bar{\alpha}_{i}^{\prime },\alpha _{i};0)  \notag \\
&&\times \rho (\bar{\alpha}_{i},\alpha _{i}^{\prime };0),  \label{timeevo}
\end{eqnarray}%
where $\rho (\bar{\alpha}_{f},\alpha _{f}^{\prime };t)\equiv \langle \bar{\alpha}_{f}|\rho\left( t\right) |\alpha_{f}^{\prime }\rangle $ and $\rho (\bar{\alpha}_{i},\alpha _{i}^{\prime };0)=\langle \bar{\alpha}_{i}|\rho(0)|\alpha _{i}^{\prime } \rangle$, with $|\alpha\rangle=e^{\alpha\hat{a}^\dag}|0\rangle$ and $d\mu(\alpha)=e^{-|\alpha|^2}d^2\alpha/\pi$. After eliminating the degrees of freedom of the reservoir, the propagation functional $\mathcal{J}(\bar{\alpha}_{f},\alpha _{f}^{\prime };t|\bar{\alpha}_{i}^{\prime },\alpha _{i};0)$ reads \begin{eqnarray}
&&\mathcal{J}(\bar{\alpha}_{f},\alpha _{f}^{\prime };t|\bar{\alpha}%
_{i}^{\prime },\alpha _{i};0)=M(t)\exp \{J_{1}(t)\bar{\alpha}_{f}\alpha _{i}
\notag \\
&&~~~+J_{2}\left( t\right) \bar{\alpha}_{f}\alpha _{f}^{\prime
}-[J_{3}\left( t\right)-1] \bar{\alpha}_{i}^{\prime }\alpha _{i}+J_{1}^{\ast }(t)%
\bar{\alpha}_{i}^{\prime }\alpha _{f}^{\prime }\},  ~~\label{prord}
\end{eqnarray}%
where
\begin{eqnarray}
M(t) &=&[1+v(t)]^{-1},~J_{1}(t)=M(t)u(t), \\
J_{2}(t) &=&M(t)v(t),~J_{3}(t)=M(t)|u(t)|^{2}.
\end{eqnarray}
The equations of motion of $u(t)$ and $v(t)$ take the forms of Eqs. \eqref{u} and \eqref{extv}, respectively.

The thermometer is initially in a coherent state
\begin{equation}
\rho ( \bar{\alpha}_{i},\alpha _{i}^{\prime };0) =\exp ( -\left\vert
\alpha_{0}\right\vert ^{2}+\bar{\alpha}_{i}\alpha _{0}+\bar{\alpha}_{0}\alpha_{i}^{\prime }).  \label{bbct}
\end{equation}%
Substituting (\ref{bbct}) into Eq. (\ref{timeevo}) and performing the Gaussian
integration, we get
\begin{eqnarray}
\rho ( \bar{\alpha}_{f},\alpha _{f}^{\prime };t)&=&M(t)\exp [ -J_{3}(
t)\left\vert \alpha _{0}\right\vert^{2}+J_{1}(t)\bar{\alpha}_{f}\alpha _{0}
\notag \\
&&+J_{2}( t) \bar{\alpha}_{f}\alpha _{f}^{\prime}+J_{1}^{\ast }(t)\bar{\alpha%
}_{0}\alpha _{f}^{\prime }].  \label{arho}
\end{eqnarray}
Remembering that the density matrix is expressed as $\rho(t)=\int
d\mu(\alpha_f)d\mu(\alpha_f^{\prime })\rho ( \bar{\alpha}_{f},\alpha
_{f}^{\prime };t)|\alpha_f\rangle\langle \bar{\alpha}_f'|$, we obtain
\begin{eqnarray}
\rho(t)&=&M(t)\exp [-J_{3}(t)\left\vert \alpha _{0}\right\vert ^{2}]\exp \left[J_{1}(t)\alpha _{0}a^{\dag }\right]\nonumber\\
&& \times\exp \left[ \ln J_{2}(t)a^{\dag }a\right] \exp \left[ J_{1}^{\ast }(t)\alpha _{0}^{\ast }a\right].\nonumber\\
&=&M(t)\hat{\mathcal D}_te^{\hat{a}^{\dag }\hat{a}\ln J_{2}(t)}\hat{\mathcal D}^{\dag }_t,\label{smrhotfd}
\end{eqnarray}
where $\hat{\mathcal{D}}_t=\exp[\alpha_0 u(t)\hat{a}^\dag-\alpha_0^*u^*(t)\hat{a}]$. The last equality of Eq. \eqref{smrhotfd} can be proven as follows.

\begin{proof}Temporally neglecting the arguments ``$(t)$'' of each time-dependent functions for brevity, we have
\begin{eqnarray}
\hat{\mathcal D}_te^{\hat{a}^{\dag }\hat{a}\ln J_{2}}\hat{\mathcal D}^{\dag }_t&=&\exp [(a^{\dag }-u^{\ast }\alpha _{0}^{\ast })(a-u\alpha _{0})\ln J_{2}]\nonumber\\
&=&J_2 ^{\left\vert u\alpha _{0}\right\vert^{2}}\exp [-\ln J_{2}u\alpha _{0}\hat{a}^{\dag }+\ln J_{2}\hat{a}^{\dag }\hat{a}\nonumber\\
&&-\ln J_{2}u^{\ast }\alpha _{0}^{\ast }\hat{a}].\label{smdmf}
\end{eqnarray}
Then according to the rule of disentangling an exponential of operators of a harmonic oscillator algebra, i.e.,
$e^{(\beta _{+}\hat{a}^{\dag }+\beta _{0}\hat{a}^{\dag }\hat{a}+\beta _{-}\hat{a})}=e^{f_{+}\hat{a}^{\dag }}e^{f_{0}\hat{a}^{\dag }\hat{a}}e^{f_{-}\hat{a}}e^{g}$,
with $f_{\pm}={\beta _{\pm}(e^{\beta _{0}}-1)}/{\beta _{0}}$,~$f_0=\beta_0$, and $g ={\beta _{+}\beta _{-}(e^{\beta _{0} }-1-\beta_0)}/{\beta_{0}^{2}}$,
we readily obtain
\begin{eqnarray}
\hat{\mathcal D}_te^{\hat{a}^{\dag }\hat{a}\ln J_{2}}\hat{\mathcal D}^{\dag }_t&=&e^{-\left\vert u\alpha _{0}\right\vert^{2}(1-J_{2})}e^{(1-J_{2})u\alpha _{0}\hat{a}^{\dag }}e^{\ln J_2\hat{a}^{\dag }\hat{a}}\nonumber\\
&&\times e^{(1-J_{2})u^{\ast }\alpha _{0}^{\ast }\hat{a}}\nonumber\\
&=&e^{-J_3\left\vert \alpha _{0}\right\vert^{2}}e^{J_1\alpha _{0}\hat{a}^{\dag }}e^{\ln J_2\hat{a}^{\dag }\hat{a}}e^{J_1^*\alpha _{0}^{\ast }\hat{a}},
\end{eqnarray}which just the first equality of Eq. \eqref{smdmf}.
\end{proof}

After inserting the complete basis $\sum_n|n\rangle\langle n|=1$ into Eq. \eqref{smdmf}, we have
\begin{eqnarray}
\rho(t)&=&M(t)\hat{\mathcal D}_t\sum_{n}e^{n\ln J_{2}(t)}|n\rangle\langle n|\hat{\mathcal D}^{\dag }_t\nonumber\\
&=&\hat{\mathcal D}_t\sum_{n}{v(t)^n\over [1+v(t)]^{n+1}}|n\rangle\langle n|\hat{\mathcal D}^{\dag }_t.
\end{eqnarray}

\section{Born-Markovian limit of $u(t)$ and $v(t)$}\label{DrBMas}
 Defining $u(t)=e^{-i\omega_0 t}u'(t)$, we rewrite Eq. \eqref{u} as
\begin{equation}
\dot{ u}'(t)+\int_0^td\tau\int_0^\infty d\omega J(\omega)e^{-i(\omega-\omega_0)(t-\tau)}u'(\tau)=0.\label{su}
\end{equation}When the thermometer-reservoir coupling is weak and the time scale of the reservoir correlation function is much smaller than that of the thermometer, we can calculate the Born-Markovian approximate solution of Eq. (\ref{su}) via neglecting the memory effect \cite{breuer2002theory}, i.e., $u'(\tau)\simeq u'(t)$, and extending the upper limit of the integral to infinity, i.e. $\int_0^td\tau\simeq\int_0^\infty d\tau$. Utilization of the identity $\lim_{t\rightarrow\infty}\int_0^t d\tau e^{-i(\omega-\omega_0)(t-\tau)}=\pi\delta(\omega-\omega_0)+i\mathcal{P}{1\over \omega_0-\omega}$, with $\mathcal{P}$ being the Cauchy principal value, results in $u'_\text{MA}(t)=e^{-[\kappa+i\Delta(\omega_0)]t}$, where $\kappa=\pi J(\omega_0)$ and $\Delta(\omega_0)=\mathcal{P}\int_0^\infty{J(\omega)\over\omega_0-\omega}d\omega$. We thus have the Born-Markovian approximate solution of $u(t)$ as $u_\text{MA}(t)=e^{-[\kappa+i(\omega_0+\Delta(\omega_0))]t}$.

Equation \eqref{extv} is rewritten as\begin{equation}
v(t)=\int_0^\infty d\omega \bar{n}(\omega)A_\omega(t),\label{smvt}
\end{equation} where $A_\omega(t)=J(\omega)|\tilde{u}_\omega(t)|^2$ with $\tilde{u}_\omega(t)=\int_0^t d\tau u(\tau)e^{i\omega\tau}$. The function $A_\omega(t)$ is further recast into\begin{equation}
\int_0^\infty d\omega A_\omega(t)=\int_0^td\tau_1\int_0^td\tau_2 u^*(\tau_2)\mu(\tau_2-\tau_1)u(\tau_1).
\end{equation} Its derivative with respect to time reads
\begin{eqnarray}
{d\over dt}\int_0^\infty d\omega A_\omega(t)&=&u(t)\Big[\int_0^t d\tau_2u(\tau_2)\mu(t-\tau_2)\Big]^*\nonumber\\
&&+u^*(t)\int_0^td\tau_1 \mu(t-\tau_1)u(\tau_1)\nonumber\\
&=&-{d|u(t)|^2\over dt},\label{smda}
\end{eqnarray}where Eq. \eqref{u} has been used in the last equality. Then integrating Eq. \eqref{smda} with time and using $\int_0^\infty d\omega A_\omega(0)=0$, we readily obtain
\begin{equation}\int_0^\infty d\omega A_\omega(t)=1-|u(t)|^2.\label{Sai}\end{equation}

Under the Born-Markovian approximation, we safely replace $\bar{n}(\omega)$ in Eq. \eqref{smvt} by $\bar{n}(\omega_0)$. Then we have
\begin{eqnarray}
v_\text{MA}(t)&=&\bar{n}(\omega_0)\int_0^\infty d\omega A_\omega(t)=\bar{n}(\omega_0)[1-|u_\text{MA}(t)|^2]\nonumber\\
&=&\bar{n}(\omega_0)(1-e^{-2\kappa t}).
\end{eqnarray}

\section{Derivation of the QFI}\label{drqfi}
To calculate the long-time QFI, $v(t)$ and $A_\omega(t)$ in Eq. \eqref{smvt} should be known. It can be proven that
\begin{equation}
A_\omega(\infty)=J(\omega)|\tilde{u}_\omega(\infty)|^2=\Theta(\omega)+{Z^2J(\omega)\over (\omega-E_b)^2}.\label{smaomeaga}
\end{equation}
\begin{proof}
 The exact solution of $u(t)$ is
\begin{equation}~\label{smut}
u(t)=Ze^{-iE_{b}t}+\int_{0}^{\infty}dE \Theta(E)e^{-iEt},
\end{equation}with $Z=[1+\int_{0}^{\infty}\frac{J(\omega)}{(E_{b}-\omega)^{2}}d\omega]^{-1}$ and $\Theta(E)=\frac{J(E)}{[E-\omega_{0}-\Delta(E)]^{2}+[\pi J(E)]^{2}}$. It induces
\begin{eqnarray}
 \tilde{u}_\omega(\infty)&=&\pi Z\delta(\omega-E_b)+i\mathcal{P}{Z\over\omega-E_b}\nonumber\\
 &&+\pi \Theta(\omega)+i\mathcal{P}\int_0^\infty dE{\Theta(E)\over \omega-E},\label{smuinft}
\end{eqnarray}where $\lim_{t\rightarrow\infty}\int_0^t d\tau e^{-i(E-\omega)\tau}=\pi\delta(\omega-E)+i\mathcal{P}{1\over \omega-E}$ has been used. Rewriting $\Theta(x)={i\over\pi}\text{Im}[{1\over E-\omega_0-\Delta(x)+i\pi J(x)}]$ with $x=\omega$ and $E$ and using the Hilbert transform \cite{book:1123190} $\text{Re}[h(x)]={1\over\pi}\mathcal{P}\int_{-\infty}^{+\infty}d\omega {\text{Im}[h(x)]\over \omega-x}$ for an analytical function $h(x)$, we recast the last two terms of Eq. \eqref{smuinft} as
\begin{equation}
\pi\Theta(\omega)+i\mathcal{P}\int_0^\infty dE{\Theta(E)\over \omega-E}={1\over E-\omega_0-\Delta(\omega)+i\pi J(\omega)}.\label{smltm}
\end{equation}It is noted that, in deriving Eq. \eqref{smltm}, we have extended the lower limit of the integral of $E$ from zero to $-\infty$ based on the fact that $J(E)=0$ in the regime $E\in (-\infty,0)$. Because of $E_b<0$, the first term of Eq. \eqref{smuinft} has no contribution to the integral in Eq. \eqref{smvt} and the Cauchy principal value in the second term of Eq. \eqref{smuinft} is equal to the function itself. Therefore, using Eq. \eqref{smltm}, we can safely convert Eq.  \eqref{smuinft} into
\begin{equation}
\tilde{u}_\omega(\infty)={iZ\over\omega-E_b}+{1\over E-\omega_0-\Delta(\omega)+i\pi J(\omega)},
\end{equation}which results in
\begin{eqnarray}
|\tilde{u}_\omega(\infty)|^2&=&{1\over [\omega-\omega_0-\Delta(\omega)]^2+[\pi J(\omega)]^2}+{Z^2\over (\omega-E_b)^2}\nonumber\\
&&+{\pi Z\Theta(\omega)\over\omega-E_b}.~~~~~~~~
\end{eqnarray}
The last term also has no contribution to the integral in Eq. \eqref{smvt} because, according to the residue theorem, the residue of its only pole $\omega=E_b$ is equal to zero due to $J(E_b)=0$. We thus finally arrive at Eq. \eqref{smaomeaga}.
\end{proof}

We are now ready to prove Eq. \eqref{marst}.

\begin{proof}
The QFI is calculated via $\mathcal{F}_{T}(t)={M(t)[\partial_T v(t)]^2/v(t)}$ as
\begin{eqnarray}
\mathcal{F}_T(t)&=&{M(t)\over T^2v(t)}\Big[\int_0^\infty d\omega \sqrt{A_\omega(t)}\beta\omega\sqrt{\bar{n}(\omega)}[1+\bar{n}(\omega)]\nonumber\\
&&\times\sqrt{\bar{n}(\omega)A_\omega(t)}\Big]^2.
\end{eqnarray}
Using Cauchy-Schwarz inequality $[\int_0^\infty d\omega x(\omega)y(\omega)]^2\leqslant\int_0^\infty d\omega x(\omega)^2\int_0^\infty d\omega y(\omega)^2 $  \cite{book:1123190}, we obtain
\begin{eqnarray}
\mathcal{F}_T(t)&\leqslant&{M(t)\over T^2}\int_0^\infty d\omega A_\omega(t)(\beta\omega)^2\bar{n}(\omega)[1+\bar{n}(\omega)]^2\nonumber\\
&=&M(t)\int_0^\infty d\omega \bar{{F}}_T(\omega)A_\omega(t)[1+\bar{n}(\omega)],
\end{eqnarray}where $\bar{{F}}_T(\omega)=(\beta\omega)^2 \bar{n}(\omega)[1+\bar{n}(\omega)]/ T^2$. The long-time QFI reads
\begin{equation}
\mathcal{F}_T(\infty)\leqslant {\int_0^\infty d\omega \bar{{F}}_T(\omega)A_\omega(\infty)[1+\bar{n}(\omega)]\over 1+v(\infty)}.\label{smFist}
\end{equation}

At the critical point of forming the bound state, $\omega_0=\eta\omega_c\gamma(s)$, $E_b=0$, $Z=0$, and $\Delta(0)=-\omega_0$ from Eq. \eqref{transeq}. We thus have
\begin{equation}
A_\omega(\infty)\big|_{\text{CP}}=\Theta(\omega)=\frac{J(\omega)}{[\omega-\omega_{0}-\Delta(\omega)]^{2}+[\pi J(\omega)]^{2}},
\end{equation}which has a sharp peak at $\omega=0$ due to $\lim_{\omega\rightarrow0}A_\omega(\infty)\big|_{\text{CP}}=\lim_{\omega\rightarrow0}1/[\pi^2J(\omega)]=\infty$. Therefore, the integral in Eq. \eqref{smFist} is dominated by $\omega=0$. According to $\lim_{\omega\rightarrow 0}\bar{{F}}_T(\omega)=T^{-2}$, we have
\begin{eqnarray}
\mathcal{F}_T(\infty)\Big|_\text{CP}&\leqslant& T^{-2}{\int_0^\infty d\omega A_\omega(\infty)[1+\bar{n}(\omega)]\over 1+v(\infty)}\nonumber\\
&=&T^{-2}{1-|u(\infty)|^2+v(\infty)\over 1+v(\infty)}=T^{-2},~~\label{smlimtQF}
\end{eqnarray}where Eqs. \eqref{smvt} and \eqref{Sai}, and $|u(\infty)|^2=Z^2=0$ at the critical point have been used.
\end{proof}

\begin{figure}[tbp]
\centering
\includegraphics[width=1 \columnwidth]{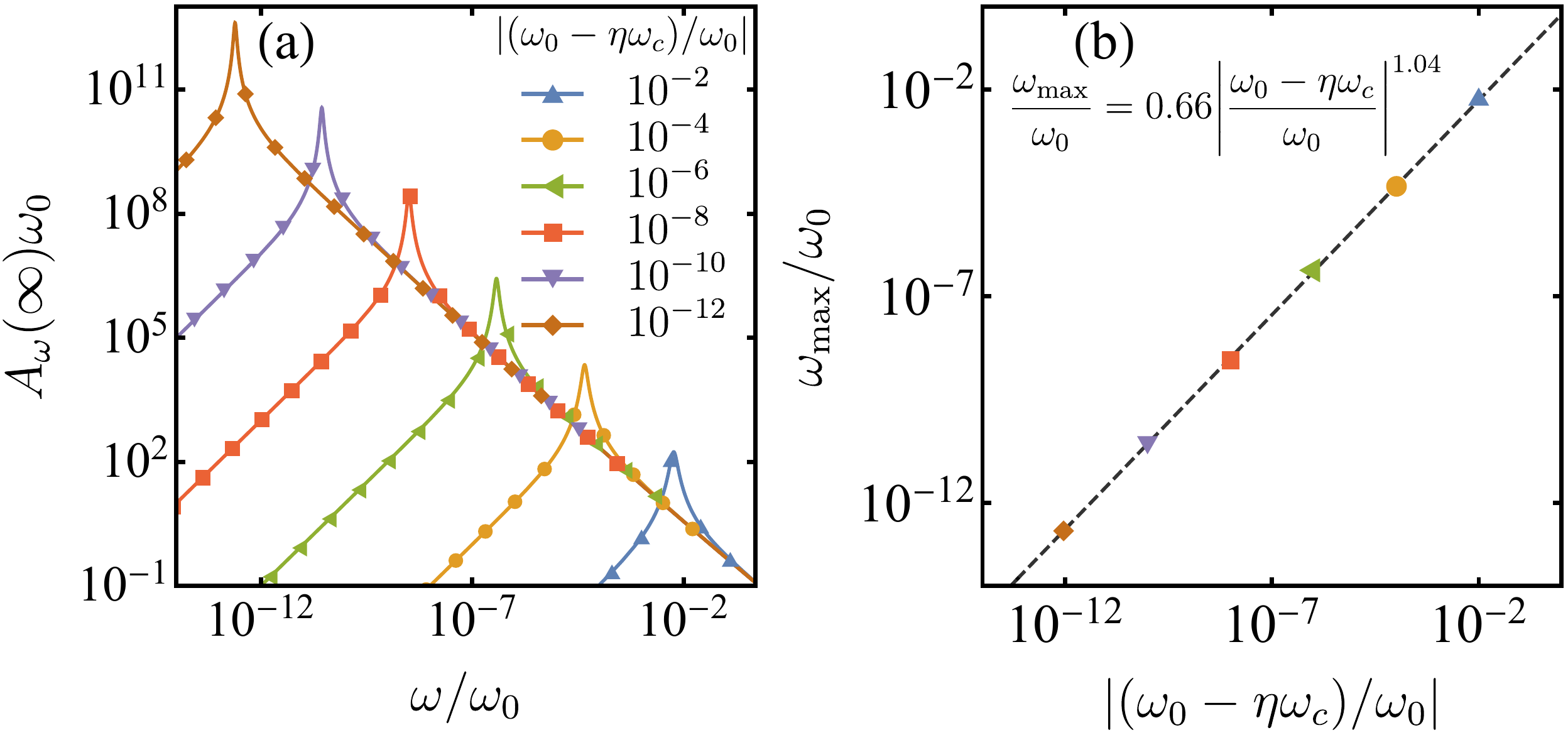}
 \caption{(a) $A_\omega(\infty)$ for different $|\omega_0-\eta\omega_c|$ for the Ohmic spectral density. (b) Numerical fitting to the frequency $\omega_\text{max}$ of the peak of $A_\omega(\infty)$ with respect to $|\omega_0-\eta\omega_c|$.  }
 \label{smfg}
\end{figure}

\section{QFI near the critical point}\label{critpndt}
The result $\mathcal{F}_T(\infty)\propto T^{-2}$ is also achievable when the parameters are near but not exactly at the critical point. Here, we give the proof of this conclusion. The function $A_\omega(\infty)$ shows a sharp peak at $\omega=0$ when the parameters are exactly at the critical point. When the parameters slightly deviate from the critical point, the sharp peak exhibits a blue shift to $\omega_\text{max}$ [see Fig. \ref{crtc}(d)]. This implies that we can approximately equate $A_\omega(\infty)$ in Eq. \eqref{smFist} as $A_\omega(\infty)\simeq\delta(\omega-\omega_\text{max})$. Then we calculate from Eq. \eqref{smFist} that
\begin{equation}
\mathcal{F}_T(\infty)\leqslant\bar{{F}}_T(\omega_\text{max})\equiv{f(\beta\omega_\text{max})\over T^2},\label{smnct}
\end{equation}
where $f(x)=x^2\bar{n}(x)[1+\bar{n}(x)]$ and Eq. \eqref{smvt} have been used. According to the fact that $f(x)$ is a monotonically decreasing function with an increase of $x$, a QFI larger than $f(1)/T^2=0.92/T^2$ is obtained as long as
\begin{equation}
\beta\omega_\text{max}\leqslant1.\label{smcdt}
\end{equation}

We now numerically explore the scaling law of $\omega_\text{max}$ near the critical point. Figure \ref{smfg}(a) shows $A_\omega(\infty)$ for different $|\omega_0-\eta\omega_c|$ near the critical point $\big|\omega_0-\eta\omega_c\big|_\text{CP}=0$ for the Ohmic spectral density. The numerical fitting in Fig. \ref{smfg}(b) reveals that the position of the peak of $A_\omega(\infty)$ scales with $|\omega_0-\eta\omega_c|$ as
\begin{equation}
 \omega_\text{max}=0.66\omega_0|(\omega_0-\eta\omega_c)/\omega_0|^{1.04}\simeq 0.66|\omega_0-\eta\omega_c|. \label{SMOMG}
\end{equation}
Substituting Eq. \eqref{SMOMG} into Eq. \eqref{smcdt}, we conclude that the QFI takes as $\mathcal{F}_T(\infty)\leqslant f(1)/T^{2}=0.92/T^2$ as long as the parameters fall in a relatively wide parameter regime near the critical point, i.e.,
\begin{equation}
|\omega_0-\eta\omega_c|\leqslant 1.52K_B T.\label{smfnr}
\end{equation}
Equation \eqref{smfnr} indicates that $\mathcal{F}_T(\infty)\leqslant 0.92/T^2$ is achievable only at single parameter point $\omega_0=\eta\omega_c$ when and only when the temperature is absolute zero. The parameter regime supporting $\mathcal{F}_T(\infty)\leqslant 0.92/T^2$ becomes wider and wider with increasing temperature. Given the fact that we never experiment at absolute zero temperature, this wide parameter regime endows our scheme with a fault tolerance to the imprecise tuning of the system parameters to reach the critical point.

\bibliography{NMT}

\begin{thebibliography}{76}%
\makeatletter
\providecommand \@ifxundefined [1]{%
 \@ifx{#1\undefined}
}%
\providecommand \@ifnum [1]{%
 \ifnum #1\expandafter \@firstoftwo
 \else \expandafter \@secondoftwo
 \fi
}%
\providecommand \@ifx [1]{%
 \ifx #1\expandafter \@firstoftwo
 \else \expandafter \@secondoftwo
 \fi
}%
\providecommand \natexlab [1]{#1}%
\providecommand \enquote  [1]{``#1''}%
\providecommand \bibnamefont  [1]{#1}%
\providecommand \bibfnamefont [1]{#1}%
\providecommand \citenamefont [1]{#1}%
\providecommand \href@noop [0]{\@secondoftwo}%
\providecommand \href [0]{\begingroup \@sanitize@url \@href}%
\providecommand \@href[1]{\@@startlink{#1}\@@href}%
\providecommand \@@href[1]{\endgroup#1\@@endlink}%
\providecommand \@sanitize@url [0]{\catcode `\\12\catcode `\$12\catcode
  `\&12\catcode `\#12\catcode `\^12\catcode `\_12\catcode `\%12\relax}%
\providecommand \@@startlink[1]{}%
\providecommand \@@endlink[0]{}%
\providecommand \url  [0]{\begingroup\@sanitize@url \@url }%
\providecommand \@url [1]{\endgroup\@href {#1}{\urlprefix }}%
\providecommand \urlprefix  [0]{URL }%
\providecommand \Eprint [0]{\href }%
\providecommand \doibase [0]{https://doi.org/}%
\providecommand \selectlanguage [0]{\@gobble}%
\providecommand \bibinfo  [0]{\@secondoftwo}%
\providecommand \bibfield  [0]{\@secondoftwo}%
\providecommand \translation [1]{[#1]}%
\providecommand \BibitemOpen [0]{}%
\providecommand \bibitemStop [0]{}%
\providecommand \bibitemNoStop [0]{.\EOS\space}%
\providecommand \EOS [0]{\spacefactor3000\relax}%
\providecommand \BibitemShut  [1]{\csname bibitem#1\endcsname}%
\let\auto@bib@innerbib\@empty
%</preamble>
\bibitem [{\citenamefont {Giazotto}\ \emph {et~al.}(2006)\citenamefont
  {Giazotto}, \citenamefont {Heikkil\"a}, \citenamefont {Luukanen},
  \citenamefont {Savin},\ and\ \citenamefont {Pekola}}]{RevModPhys.78.217}%
  \BibitemOpen
  \bibfield  {author} {\bibinfo {author} {\bibfnamefont {F.}~\bibnamefont
  {Giazotto}}, \bibinfo {author} {\bibfnamefont {T.~T.}\ \bibnamefont
  {Heikkil\"a}}, \bibinfo {author} {\bibfnamefont {A.}~\bibnamefont
  {Luukanen}}, \bibinfo {author} {\bibfnamefont {A.~M.}\ \bibnamefont
  {Savin}},\ and\ \bibinfo {author} {\bibfnamefont {J.~P.}\ \bibnamefont
  {Pekola}},\ }\bibfield  {title} {\bibinfo {title} {Opportunities for
  mesoscopics in thermometry and refrigeration: Physics and applications},\
  }\href {https://doi.org/10.1103/RevModPhys.78.217} {\bibfield  {journal}
  {\bibinfo  {journal} {Rev. Mod. Phys.}\ }\textbf {\bibinfo {volume} {78}},\
  \bibinfo {pages} {217} (\bibinfo {year} {2006})}\BibitemShut {NoStop}%
\bibitem [{\citenamefont {Carlos}\ and\ \citenamefont
  {Palacio}(2016)}]{10.1039/9781782622031}%
  \BibitemOpen
  \bibinfo {editor} {\bibfnamefont {L.~D.}\ \bibnamefont {Carlos}}\ and\
  \bibinfo {editor} {\bibfnamefont {F.}~\bibnamefont {Palacio}},\ eds.,\ \href
  {http://dx.doi.org/10.1039/9781782622031} {\emph {\bibinfo {title}
  {Thermometry at the Nanoscale}}}\ (\bibinfo  {publisher} {The Royal Society
  of Chemistry, Cambridge},\ \bibinfo {year} {2016})\BibitemShut {NoStop}%
\bibitem [{\citenamefont {De~Pasquale}\ and\ \citenamefont
  {Stace}(2018)}]{DePasquale2018}%
  \BibitemOpen
  \bibfield  {author} {\bibinfo {author} {\bibfnamefont {A.}~\bibnamefont
  {De~Pasquale}}\ and\ \bibinfo {author} {\bibfnamefont {T.~M.}\ \bibnamefont
  {Stace}},\ }in\ \href {https://doi.org/10.1007/978-3-319-99046-0_21} {\emph
  {\bibinfo {booktitle} {Thermodynamics in the Quantum Regime: Fundamental
  Aspects and New Directions}}},\ \bibinfo {editor} {edited by\ \bibinfo
  {editor} {\bibfnamefont {F.}~\bibnamefont {Binder}}, \bibinfo {editor}
  {\bibfnamefont {L.~A.}\ \bibnamefont {Correa}}, \bibinfo {editor}
  {\bibfnamefont {C.}~\bibnamefont {Gogolin}}, \bibinfo {editor} {\bibfnamefont
  {J.}~\bibnamefont {Anders}},\ and\ \bibinfo {editor} {\bibfnamefont
  {G.}~\bibnamefont {Adesso}}}\ (\bibinfo  {publisher} {Springer International
  Publishing},\ \bibinfo {address} {Cham},\ \bibinfo {year} {2018})\ p.\
  \bibinfo {pages} {503}\BibitemShut {NoStop}%
\bibitem [{\citenamefont {Mehboudi}\ \emph
  {et~al.}(2019{\natexlab{a}})\citenamefont {Mehboudi}, \citenamefont
  {Sanpera},\ and\ \citenamefont {Correa}}]{Mehboudi2019}%
  \BibitemOpen
  \bibfield  {author} {\bibinfo {author} {\bibfnamefont {M.}~\bibnamefont
  {Mehboudi}}, \bibinfo {author} {\bibfnamefont {A.}~\bibnamefont {Sanpera}},\
  and\ \bibinfo {author} {\bibfnamefont {L.~A.}\ \bibnamefont {Correa}},\
  }\bibfield  {title} {\bibinfo {title} {Thermometry in the quantum regime:
  recent theoretical progress},\ }\href
  {https://doi.org/10.1088/1751-8121/ab2828} {\bibfield  {journal} {\bibinfo
  {journal} {J. Phys. A: Math. Theor.}\ }\textbf {\bibinfo {volume} {52}},\
  \bibinfo {pages} {303001} (\bibinfo {year} {2019}{\natexlab{a}})}\BibitemShut
  {NoStop}%
\bibitem [{\citenamefont {Campisi}\ \emph {et~al.}(2011)\citenamefont
  {Campisi}, \citenamefont {H\"anggi},\ and\ \citenamefont
  {Talkner}}]{Campisi2011}%
  \BibitemOpen
  \bibfield  {author} {\bibinfo {author} {\bibfnamefont {M.}~\bibnamefont
  {Campisi}}, \bibinfo {author} {\bibfnamefont {P.}~\bibnamefont {H\"anggi}},\
  and\ \bibinfo {author} {\bibfnamefont {P.}~\bibnamefont {Talkner}},\
  }\bibfield  {title} {\bibinfo {title} {Colloquium: Quantum fluctuation
  relations: Foundations and applications},\ }\href
  {https://doi.org/10.1103/RevModPhys.83.771} {\bibfield  {journal} {\bibinfo
  {journal} {Rev. Mod. Phys.}\ }\textbf {\bibinfo {volume} {83}},\ \bibinfo
  {pages} {771} (\bibinfo {year} {2011})}\BibitemShut {NoStop}%
\bibitem [{\citenamefont {Brand{\~a}o}\ \emph {et~al.}(2015)\citenamefont
  {Brand{\~a}o}, \citenamefont {Horodecki}, \citenamefont {Ng}, \citenamefont
  {Oppenheim},\ and\ \citenamefont {Wehner}}]{Brandao2015}%
  \BibitemOpen
  \bibfield  {author} {\bibinfo {author} {\bibfnamefont {F.}~\bibnamefont
  {Brand{\~a}o}}, \bibinfo {author} {\bibfnamefont {M.}~\bibnamefont
  {Horodecki}}, \bibinfo {author} {\bibfnamefont {N.}~\bibnamefont {Ng}},
  \bibinfo {author} {\bibfnamefont {J.}~\bibnamefont {Oppenheim}},\ and\
  \bibinfo {author} {\bibfnamefont {S.}~\bibnamefont {Wehner}},\ }\bibfield
  {title} {\bibinfo {title} {The second laws of quantum thermodynamics},\
  }\href {https://doi.org/10.1073/pnas.1411728112} {\bibfield  {journal}
  {\bibinfo  {journal} {Proc. Natl. Acad. Sci. USA}\ }\textbf {\bibinfo
  {volume} {112}},\ \bibinfo {pages} {3275} (\bibinfo {year}
  {2015})}\BibitemShut {NoStop}%
\bibitem [{\citenamefont {Vinjanampathy}\ and\ \citenamefont
  {Anders}(2016)}]{doi:10.1080/00107514.2016.1201896}%
  \BibitemOpen
  \bibfield  {author} {\bibinfo {author} {\bibfnamefont {S.}~\bibnamefont
  {Vinjanampathy}}\ and\ \bibinfo {author} {\bibfnamefont {J.}~\bibnamefont
  {Anders}},\ }\bibfield  {title} {\bibinfo {title} {Quantum thermodynamics},\
  }\href {https://doi.org/10.1080/00107514.2016.1201896} {\bibfield  {journal}
  {\bibinfo  {journal} {Contemp. Phys.}\ }\textbf {\bibinfo {volume} {57}},\
  \bibinfo {pages} {545} (\bibinfo {year} {2016})}\BibitemShut {NoStop}%
\bibitem [{\citenamefont {Deffner}\ and\ \citenamefont
  {Campbell}(2019)}]{10.1088/2053-2571/ab21c6}%
  \BibitemOpen
  \bibfield  {author} {\bibinfo {author} {\bibfnamefont {S.}~\bibnamefont
  {Deffner}}\ and\ \bibinfo {author} {\bibfnamefont {S.}~\bibnamefont
  {Campbell}},\ }\href {https://doi.org/10.1088/2053-2571/ab21c6} {\emph
  {\bibinfo {title} {Quantum Thermodynamics}}}\ (\bibinfo  {publisher} {Morgan
  \& Claypool Publishers},\ \bibinfo {year} {2019})\BibitemShut {NoStop}%
\bibitem [{\citenamefont {Marzolino}\ and\ \citenamefont
  {Braun}(2013)}]{PhysRevA.88.063609}%
  \BibitemOpen
  \bibfield  {author} {\bibinfo {author} {\bibfnamefont {U.}~\bibnamefont
  {Marzolino}}\ and\ \bibinfo {author} {\bibfnamefont {D.}~\bibnamefont
  {Braun}},\ }\bibfield  {title} {\bibinfo {title} {Precision measurements of
  temperature and chemical potential of quantum gases},\ }\href
  {https://doi.org/10.1103/PhysRevA.88.063609} {\bibfield  {journal} {\bibinfo
  {journal} {Phys. Rev. A}\ }\textbf {\bibinfo {volume} {88}},\ \bibinfo
  {pages} {063609} (\bibinfo {year} {2013})}\BibitemShut {NoStop}%
\bibitem [{\citenamefont {Olf}\ \emph {et~al.}(2015)\citenamefont {Olf},
  \citenamefont {Fang}, \citenamefont {Marti}, \citenamefont {MacRae},\ and\
  \citenamefont {Stamper-Kurn}}]{OLF2015}%
  \BibitemOpen
  \bibfield  {author} {\bibinfo {author} {\bibfnamefont {R.}~\bibnamefont
  {Olf}}, \bibinfo {author} {\bibfnamefont {F.}~\bibnamefont {Fang}}, \bibinfo
  {author} {\bibfnamefont {G.~E.}\ \bibnamefont {Marti}}, \bibinfo {author}
  {\bibfnamefont {A.}~\bibnamefont {MacRae}},\ and\ \bibinfo {author}
  {\bibfnamefont {D.~M.}\ \bibnamefont {Stamper-Kurn}},\ }\bibfield  {title}
  {\bibinfo {title} {Thermometry and cooling of a bose gas to 0.02 times the
  condensation temperature},\ }\href {https://doi.org/10.1038/nphys3408}
  {\bibfield  {journal} {\bibinfo  {journal} {Nat. Phys.}\ }\textbf {\bibinfo
  {volume} {11}},\ \bibinfo {pages} {720} (\bibinfo {year} {2015})}\BibitemShut
  {NoStop}%
\bibitem [{\citenamefont {Mehboudi}\ \emph
  {et~al.}(2019{\natexlab{b}})\citenamefont {Mehboudi}, \citenamefont {Lampo},
  \citenamefont {Charalambous}, \citenamefont {Correa}, \citenamefont
  {Garc\'{\i}a-March},\ and\ \citenamefont {Lewenstein}}]{Mehboudi2019a}%
  \BibitemOpen
  \bibfield  {author} {\bibinfo {author} {\bibfnamefont {M.}~\bibnamefont
  {Mehboudi}}, \bibinfo {author} {\bibfnamefont {A.}~\bibnamefont {Lampo}},
  \bibinfo {author} {\bibfnamefont {C.}~\bibnamefont {Charalambous}}, \bibinfo
  {author} {\bibfnamefont {L.~A.}\ \bibnamefont {Correa}}, \bibinfo {author}
  {\bibfnamefont {M.~A.}\ \bibnamefont {Garc\'{\i}a-March}},\ and\ \bibinfo
  {author} {\bibfnamefont {M.}~\bibnamefont {Lewenstein}},\ }\bibfield  {title}
  {\bibinfo {title} {Using polarons for sub-n{K} quantum nondemolition
  thermometry in a {B}ose-{E}instein condensate},\ }\href
  {https://doi.org/10.1103/PhysRevLett.122.030403} {\bibfield  {journal}
  {\bibinfo  {journal} {Phys. Rev. Lett.}\ }\textbf {\bibinfo {volume} {122}},\
  \bibinfo {pages} {030403} (\bibinfo {year} {2019}{\natexlab{b}})}\BibitemShut
  {NoStop}%
\bibitem [{\citenamefont {Bouton}\ \emph {et~al.}(2020)\citenamefont {Bouton},
  \citenamefont {Nettersheim}, \citenamefont {Adam}, \citenamefont {Schmidt},
  \citenamefont {Mayer}, \citenamefont {Lausch}, \citenamefont {Tiemann},\ and\
  \citenamefont {Widera}}]{PhysRevX.10.011018}%
  \BibitemOpen
  \bibfield  {author} {\bibinfo {author} {\bibfnamefont {Q.}~\bibnamefont
  {Bouton}}, \bibinfo {author} {\bibfnamefont {J.}~\bibnamefont {Nettersheim}},
  \bibinfo {author} {\bibfnamefont {D.}~\bibnamefont {Adam}}, \bibinfo {author}
  {\bibfnamefont {F.}~\bibnamefont {Schmidt}}, \bibinfo {author} {\bibfnamefont
  {D.}~\bibnamefont {Mayer}}, \bibinfo {author} {\bibfnamefont
  {T.}~\bibnamefont {Lausch}}, \bibinfo {author} {\bibfnamefont
  {E.}~\bibnamefont {Tiemann}},\ and\ \bibinfo {author} {\bibfnamefont
  {A.}~\bibnamefont {Widera}},\ }\bibfield  {title} {\bibinfo {title}
  {Single-atom quantum probes for ultracold gases boosted by nonequilibrium
  spin dynamics},\ }\href {https://doi.org/10.1103/PhysRevX.10.011018}
  {\bibfield  {journal} {\bibinfo  {journal} {Phys. Rev. X}\ }\textbf {\bibinfo
  {volume} {10}},\ \bibinfo {pages} {011018} (\bibinfo {year}
  {2020})}\BibitemShut {NoStop}%
\bibitem [{\citenamefont {Mitchison}\ \emph {et~al.}(2020)\citenamefont
  {Mitchison}, \citenamefont {Fogarty}, \citenamefont {Guarnieri},
  \citenamefont {Campbell}, \citenamefont {Busch},\ and\ \citenamefont
  {Goold}}]{PhysRevLett.125.080402}%
  \BibitemOpen
  \bibfield  {author} {\bibinfo {author} {\bibfnamefont {M.~T.}\ \bibnamefont
  {Mitchison}}, \bibinfo {author} {\bibfnamefont {T.}~\bibnamefont {Fogarty}},
  \bibinfo {author} {\bibfnamefont {G.}~\bibnamefont {Guarnieri}}, \bibinfo
  {author} {\bibfnamefont {S.}~\bibnamefont {Campbell}}, \bibinfo {author}
  {\bibfnamefont {T.}~\bibnamefont {Busch}},\ and\ \bibinfo {author}
  {\bibfnamefont {J.}~\bibnamefont {Goold}},\ }\bibfield  {title} {\bibinfo
  {title} {In situ thermometry of a cold fermi gas via dephasing impurities},\
  }\href {https://doi.org/10.1103/PhysRevLett.125.080402} {\bibfield  {journal}
  {\bibinfo  {journal} {Phys. Rev. Lett.}\ }\textbf {\bibinfo {volume} {125}},\
  \bibinfo {pages} {080402} (\bibinfo {year} {2020})}\BibitemShut {NoStop}%
\bibitem [{\citenamefont {Stace}(2010)}]{PhysRevA.82.011611}%
  \BibitemOpen
  \bibfield  {author} {\bibinfo {author} {\bibfnamefont {T.~M.}\ \bibnamefont
  {Stace}},\ }\bibfield  {title} {\bibinfo {title} {Quantum limits of
  thermometry},\ }\href {https://doi.org/10.1103/PhysRevA.82.011611} {\bibfield
   {journal} {\bibinfo  {journal} {Phys. Rev. A}\ }\textbf {\bibinfo {volume}
  {82}},\ \bibinfo {pages} {011611(R)} (\bibinfo {year} {2010})}\BibitemShut
  {NoStop}%
\bibitem [{\citenamefont {Brunelli}\ \emph {et~al.}(2011)\citenamefont
  {Brunelli}, \citenamefont {Olivares},\ and\ \citenamefont
  {Paris}}]{Brunelli2011}%
  \BibitemOpen
  \bibfield  {author} {\bibinfo {author} {\bibfnamefont {M.}~\bibnamefont
  {Brunelli}}, \bibinfo {author} {\bibfnamefont {S.}~\bibnamefont {Olivares}},\
  and\ \bibinfo {author} {\bibfnamefont {M.~G.~A.}\ \bibnamefont {Paris}},\
  }\bibfield  {title} {\bibinfo {title} {Qubit thermometry for micromechanical
  resonators},\ }\href {https://doi.org/10.1103/PhysRevA.84.032105} {\bibfield
  {journal} {\bibinfo  {journal} {Phys. Rev. A}\ }\textbf {\bibinfo {volume}
  {84}},\ \bibinfo {pages} {032105} (\bibinfo {year} {2011})}\BibitemShut
  {NoStop}%
\bibitem [{\citenamefont {Jevtic}\ \emph {et~al.}(2015)\citenamefont {Jevtic},
  \citenamefont {Newman}, \citenamefont {Rudolph},\ and\ \citenamefont
  {Stace}}]{Jevtic2015}%
  \BibitemOpen
  \bibfield  {author} {\bibinfo {author} {\bibfnamefont {S.}~\bibnamefont
  {Jevtic}}, \bibinfo {author} {\bibfnamefont {D.}~\bibnamefont {Newman}},
  \bibinfo {author} {\bibfnamefont {T.}~\bibnamefont {Rudolph}},\ and\ \bibinfo
  {author} {\bibfnamefont {T.~M.}\ \bibnamefont {Stace}},\ }\bibfield  {title}
  {\bibinfo {title} {Single-qubit thermometry},\ }\href
  {https://doi.org/10.1103/PhysRevA.91.012331} {\bibfield  {journal} {\bibinfo
  {journal} {Phys. Rev. A}\ }\textbf {\bibinfo {volume} {91}},\ \bibinfo
  {pages} {012331} (\bibinfo {year} {2015})}\BibitemShut {NoStop}%
\bibitem [{\citenamefont {Correa}\ \emph {et~al.}(2015)\citenamefont {Correa},
  \citenamefont {Mehboudi}, \citenamefont {Adesso},\ and\ \citenamefont
  {Sanpera}}]{Correa2015}%
  \BibitemOpen
  \bibfield  {author} {\bibinfo {author} {\bibfnamefont {L.~A.}\ \bibnamefont
  {Correa}}, \bibinfo {author} {\bibfnamefont {M.}~\bibnamefont {Mehboudi}},
  \bibinfo {author} {\bibfnamefont {G.}~\bibnamefont {Adesso}},\ and\ \bibinfo
  {author} {\bibfnamefont {A.}~\bibnamefont {Sanpera}},\ }\bibfield  {title}
  {\bibinfo {title} {Individual quantum probes for optimal thermometry},\
  }\href {https://doi.org/10.1103/PhysRevLett.114.220405} {\bibfield  {journal}
  {\bibinfo  {journal} {Phys. Rev. Lett.}\ }\textbf {\bibinfo {volume} {114}},\
  \bibinfo {pages} {220405} (\bibinfo {year} {2015})}\BibitemShut {NoStop}%
\bibitem [{\citenamefont {Hofer}\ \emph {et~al.}(2017)\citenamefont {Hofer},
  \citenamefont {Brask}, \citenamefont {Perarnau-Llobet},\ and\ \citenamefont
  {Brunner}}]{Hofer2017}%
  \BibitemOpen
  \bibfield  {author} {\bibinfo {author} {\bibfnamefont {P.~P.}\ \bibnamefont
  {Hofer}}, \bibinfo {author} {\bibfnamefont {J.~B.}\ \bibnamefont {Brask}},
  \bibinfo {author} {\bibfnamefont {M.}~\bibnamefont {Perarnau-Llobet}},\ and\
  \bibinfo {author} {\bibfnamefont {N.}~\bibnamefont {Brunner}},\ }\bibfield
  {title} {\bibinfo {title} {Quantum thermal machine as a thermometer},\ }\href
  {https://doi.org/10.1103/PhysRevLett.119.090603} {\bibfield  {journal}
  {\bibinfo  {journal} {Phys. Rev. Lett.}\ }\textbf {\bibinfo {volume} {119}},\
  \bibinfo {pages} {090603} (\bibinfo {year} {2017})}\BibitemShut {NoStop}%
\bibitem [{\citenamefont {Correa}\ \emph {et~al.}(2017)\citenamefont {Correa},
  \citenamefont {Perarnau-Llobet}, \citenamefont {Hovhannisyan}, \citenamefont
  {Hern\'andez-Santana}, \citenamefont {Mehboudi},\ and\ \citenamefont
  {Sanpera}}]{Correa2017}%
  \BibitemOpen
  \bibfield  {author} {\bibinfo {author} {\bibfnamefont {L.~A.}\ \bibnamefont
  {Correa}}, \bibinfo {author} {\bibfnamefont {M.}~\bibnamefont
  {Perarnau-Llobet}}, \bibinfo {author} {\bibfnamefont {K.~V.}\ \bibnamefont
  {Hovhannisyan}}, \bibinfo {author} {\bibfnamefont {S.}~\bibnamefont
  {Hern\'andez-Santana}}, \bibinfo {author} {\bibfnamefont {M.}~\bibnamefont
  {Mehboudi}},\ and\ \bibinfo {author} {\bibfnamefont {A.}~\bibnamefont
  {Sanpera}},\ }\bibfield  {title} {\bibinfo {title} {Enhancement of
  low-temperature thermometry by strong coupling},\ }\href
  {https://doi.org/10.1103/PhysRevA.96.062103} {\bibfield  {journal} {\bibinfo
  {journal} {Phys. Rev. A}\ }\textbf {\bibinfo {volume} {96}},\ \bibinfo
  {pages} {062103} (\bibinfo {year} {2017})}\BibitemShut {NoStop}%
\bibitem [{\citenamefont {Campbell}\ \emph {et~al.}(2017)\citenamefont
  {Campbell}, \citenamefont {Mehboudi}, \citenamefont {Chiara},\ and\
  \citenamefont {Paternostro}}]{Campbell_2017}%
  \BibitemOpen
  \bibfield  {author} {\bibinfo {author} {\bibfnamefont {S.}~\bibnamefont
  {Campbell}}, \bibinfo {author} {\bibfnamefont {M.}~\bibnamefont {Mehboudi}},
  \bibinfo {author} {\bibfnamefont {G.~D.}\ \bibnamefont {Chiara}},\ and\
  \bibinfo {author} {\bibfnamefont {M.}~\bibnamefont {Paternostro}},\
  }\bibfield  {title} {\bibinfo {title} {Global and local thermometry schemes
  in coupled quantum systems},\ }\href
  {https://doi.org/10.1088/1367-2630/aa7fac} {\bibfield  {journal} {\bibinfo
  {journal} {New J. Phys.}\ }\textbf {\bibinfo {volume} {19}},\ \bibinfo
  {pages} {103003} (\bibinfo {year} {2017})}\BibitemShut {NoStop}%
\bibitem [{\citenamefont {Kiilerich}\ \emph {et~al.}(2018)\citenamefont
  {Kiilerich}, \citenamefont {De~Pasquale},\ and\ \citenamefont
  {Giovannetti}}]{Kiilerich2018}%
  \BibitemOpen
  \bibfield  {author} {\bibinfo {author} {\bibfnamefont {A.~H.}\ \bibnamefont
  {Kiilerich}}, \bibinfo {author} {\bibfnamefont {A.}~\bibnamefont
  {De~Pasquale}},\ and\ \bibinfo {author} {\bibfnamefont {V.}~\bibnamefont
  {Giovannetti}},\ }\bibfield  {title} {\bibinfo {title} {Dynamical approach to
  ancilla-assisted quantum thermometry},\ }\href
  {https://doi.org/10.1103/PhysRevA.98.042124} {\bibfield  {journal} {\bibinfo
  {journal} {Phys. Rev. A}\ }\textbf {\bibinfo {volume} {98}},\ \bibinfo
  {pages} {042124} (\bibinfo {year} {2018})}\BibitemShut {NoStop}%
\bibitem [{\citenamefont {Feyles}\ \emph {et~al.}(2019)\citenamefont {Feyles},
  \citenamefont {Mancino}, \citenamefont {Sbroscia}, \citenamefont {Gianani},\
  and\ \citenamefont {Barbieri}}]{Feyles2019}%
  \BibitemOpen
  \bibfield  {author} {\bibinfo {author} {\bibfnamefont {M.~M.}\ \bibnamefont
  {Feyles}}, \bibinfo {author} {\bibfnamefont {L.}~\bibnamefont {Mancino}},
  \bibinfo {author} {\bibfnamefont {M.}~\bibnamefont {Sbroscia}}, \bibinfo
  {author} {\bibfnamefont {I.}~\bibnamefont {Gianani}},\ and\ \bibinfo {author}
  {\bibfnamefont {M.}~\bibnamefont {Barbieri}},\ }\bibfield  {title} {\bibinfo
  {title} {Dynamical role of quantum signatures in quantum thermometry},\
  }\href {https://doi.org/10.1103/PhysRevA.99.062114} {\bibfield  {journal}
  {\bibinfo  {journal} {Phys. Rev. A}\ }\textbf {\bibinfo {volume} {99}},\
  \bibinfo {pages} {062114} (\bibinfo {year} {2019})}\BibitemShut {NoStop}%
\bibitem [{\citenamefont {Mukherjee}\ \emph {et~al.}(2019)\citenamefont
  {Mukherjee}, \citenamefont {Zwick}, \citenamefont {Ghosh}, \citenamefont
  {Chen},\ and\ \citenamefont {Kurizki}}]{Mukherjee2019}%
  \BibitemOpen
  \bibfield  {author} {\bibinfo {author} {\bibfnamefont {V.}~\bibnamefont
  {Mukherjee}}, \bibinfo {author} {\bibfnamefont {A.}~\bibnamefont {Zwick}},
  \bibinfo {author} {\bibfnamefont {A.}~\bibnamefont {Ghosh}}, \bibinfo
  {author} {\bibfnamefont {X.}~\bibnamefont {Chen}},\ and\ \bibinfo {author}
  {\bibfnamefont {G.}~\bibnamefont {Kurizki}},\ }\bibfield  {title} {\bibinfo
  {title} {Enhanced precision bound of low-temperature quantum thermometry via
  dynamical control},\ }\href {https://doi.org/10.1038/s42005-019-0265-y}
  {\bibfield  {journal} {\bibinfo  {journal} {Commun. Phys.}\ }\textbf
  {\bibinfo {volume} {2}},\ \bibinfo {pages} {162} (\bibinfo {year}
  {2019})}\BibitemShut {NoStop}%
\bibitem [{\citenamefont {Potts}\ \emph {et~al.}(2019)\citenamefont {Potts},
  \citenamefont {Brask},\ and\ \citenamefont
  {Brunner}}]{Potts2019fundamentallimits}%
  \BibitemOpen
  \bibfield  {author} {\bibinfo {author} {\bibfnamefont {P.~P.}\ \bibnamefont
  {Potts}}, \bibinfo {author} {\bibfnamefont {J.~B.}\ \bibnamefont {Brask}},\
  and\ \bibinfo {author} {\bibfnamefont {N.}~\bibnamefont {Brunner}},\
  }\bibfield  {title} {\bibinfo {title} {Fundamental limits on low-temperature
  quantum thermometry with finite resolution},\ }\href
  {https://doi.org/10.22331/q-2019-07-09-161} {\bibfield  {journal} {\bibinfo
  {journal} {{Quantum}}\ }\textbf {\bibinfo {volume} {3}},\ \bibinfo {pages}
  {161} (\bibinfo {year} {2019})}\BibitemShut {NoStop}%
\bibitem [{\citenamefont {J\o{}rgensen}\ \emph {et~al.}(2020)\citenamefont
  {J\o{}rgensen}, \citenamefont {Potts}, \citenamefont {Paris},\ and\
  \citenamefont {Brask}}]{PhysRevResearch.2.033394}%
  \BibitemOpen
  \bibfield  {author} {\bibinfo {author} {\bibfnamefont {M.~R.}\ \bibnamefont
  {J\o{}rgensen}}, \bibinfo {author} {\bibfnamefont {P.~P.}\ \bibnamefont
  {Potts}}, \bibinfo {author} {\bibfnamefont {M.~G.~A.}\ \bibnamefont
  {Paris}},\ and\ \bibinfo {author} {\bibfnamefont {J.~B.}\ \bibnamefont
  {Brask}},\ }\bibfield  {title} {\bibinfo {title} {Tight bound on
  finite-resolution quantum thermometry at low temperatures},\ }\href
  {https://doi.org/10.1103/PhysRevResearch.2.033394} {\bibfield  {journal}
  {\bibinfo  {journal} {Phys. Rev. Research}\ }\textbf {\bibinfo {volume}
  {2}},\ \bibinfo {pages} {033394} (\bibinfo {year} {2020})}\BibitemShut
  {NoStop}%
\bibitem [{\citenamefont {Montenegro}\ \emph {et~al.}(2020)\citenamefont
  {Montenegro}, \citenamefont {Genoni}, \citenamefont {Bayat},\ and\
  \citenamefont {Paris}}]{Montenegro2020}%
  \BibitemOpen
  \bibfield  {author} {\bibinfo {author} {\bibfnamefont {V.}~\bibnamefont
  {Montenegro}}, \bibinfo {author} {\bibfnamefont {M.~G.}\ \bibnamefont
  {Genoni}}, \bibinfo {author} {\bibfnamefont {A.}~\bibnamefont {Bayat}},\ and\
  \bibinfo {author} {\bibfnamefont {M.~G.~A.}\ \bibnamefont {Paris}},\
  }\bibfield  {title} {\bibinfo {title} {Mechanical oscillator thermometry in
  the nonlinear optomechanical regime},\ }\href
  {https://doi.org/10.1103/PhysRevResearch.2.043338} {\bibfield  {journal}
  {\bibinfo  {journal} {Phys. Rev. Research}\ }\textbf {\bibinfo {volume}
  {2}},\ \bibinfo {pages} {043338} (\bibinfo {year} {2020})}\BibitemShut
  {NoStop}%
\bibitem [{\citenamefont {Gebbia}\ \emph {et~al.}(2020)\citenamefont {Gebbia},
  \citenamefont {Benedetti}, \citenamefont {Benatti}, \citenamefont
  {Floreanini}, \citenamefont {Bina},\ and\ \citenamefont
  {Paris}}]{Gebbia2020}%
  \BibitemOpen
  \bibfield  {author} {\bibinfo {author} {\bibfnamefont {F.}~\bibnamefont
  {Gebbia}}, \bibinfo {author} {\bibfnamefont {C.}~\bibnamefont {Benedetti}},
  \bibinfo {author} {\bibfnamefont {F.}~\bibnamefont {Benatti}}, \bibinfo
  {author} {\bibfnamefont {R.}~\bibnamefont {Floreanini}}, \bibinfo {author}
  {\bibfnamefont {M.}~\bibnamefont {Bina}},\ and\ \bibinfo {author}
  {\bibfnamefont {M.~G.~A.}\ \bibnamefont {Paris}},\ }\bibfield  {title}
  {\bibinfo {title} {Two-qubit quantum probes for the temperature of an ohmic
  environment},\ }\href {https://doi.org/10.1103/PhysRevA.101.032112}
  {\bibfield  {journal} {\bibinfo  {journal} {Phys. Rev. A}\ }\textbf {\bibinfo
  {volume} {101}},\ \bibinfo {pages} {032112} (\bibinfo {year}
  {2020})}\BibitemShut {NoStop}%
\bibitem [{\citenamefont {Planella}\ \emph {et~al.}(2021)\citenamefont
  {Planella}, \citenamefont {Cenni}, \citenamefont {Acin},\ and\ \citenamefont
  {Mehboudi}}]{planella2021bathinduced}%
  \BibitemOpen
  \bibfield  {author} {\bibinfo {author} {\bibfnamefont {G.}~\bibnamefont
  {Planella}}, \bibinfo {author} {\bibfnamefont {M.~F.~B.}\ \bibnamefont
  {Cenni}}, \bibinfo {author} {\bibfnamefont {A.}~\bibnamefont {Acin}},\ and\
  \bibinfo {author} {\bibfnamefont {M.}~\bibnamefont {Mehboudi}},\ }\href@noop
  {} {\bibinfo {title} {Bath-induced correlations lead to sub-shot-noise
  thermometry precision}} (\bibinfo {year} {2021}),\ \Eprint
  {https://arxiv.org/abs/2001.11812} {arXiv:2001.11812} \BibitemShut {NoStop}%
\bibitem [{\citenamefont {Guarnieri}\ \emph {et~al.}(2019)\citenamefont
  {Guarnieri}, \citenamefont {Landi}, \citenamefont {Clark},\ and\
  \citenamefont {Goold}}]{PhysRevResearch.1.033021}%
  \BibitemOpen
  \bibfield  {author} {\bibinfo {author} {\bibfnamefont {G.}~\bibnamefont
  {Guarnieri}}, \bibinfo {author} {\bibfnamefont {G.~T.}\ \bibnamefont
  {Landi}}, \bibinfo {author} {\bibfnamefont {S.~R.}\ \bibnamefont {Clark}},\
  and\ \bibinfo {author} {\bibfnamefont {J.}~\bibnamefont {Goold}},\ }\bibfield
   {title} {\bibinfo {title} {Thermodynamics of precision in quantum
  nonequilibrium steady states},\ }\href
  {https://doi.org/10.1103/PhysRevResearch.1.033021} {\bibfield  {journal}
  {\bibinfo  {journal} {Phys. Rev. Research}\ }\textbf {\bibinfo {volume}
  {1}},\ \bibinfo {pages} {033021} (\bibinfo {year} {2019})}\BibitemShut
  {NoStop}%
\bibitem [{\citenamefont {De~Pasquale}\ \emph {et~al.}(2017)\citenamefont
  {De~Pasquale}, \citenamefont {Yuasa},\ and\ \citenamefont
  {Giovannetti}}]{DePasquale2017}%
  \BibitemOpen
  \bibfield  {author} {\bibinfo {author} {\bibfnamefont {A.}~\bibnamefont
  {De~Pasquale}}, \bibinfo {author} {\bibfnamefont {K.}~\bibnamefont {Yuasa}},\
  and\ \bibinfo {author} {\bibfnamefont {V.}~\bibnamefont {Giovannetti}},\
  }\bibfield  {title} {\bibinfo {title} {Estimating temperature via sequential
  measurements},\ }\href {https://doi.org/10.1103/PhysRevA.96.012316}
  {\bibfield  {journal} {\bibinfo  {journal} {Phys. Rev. A}\ }\textbf {\bibinfo
  {volume} {96}},\ \bibinfo {pages} {012316} (\bibinfo {year}
  {2017})}\BibitemShut {NoStop}%
\bibitem [{\citenamefont {Cavina}\ \emph {et~al.}(2018)\citenamefont {Cavina},
  \citenamefont {Mancino}, \citenamefont {De~Pasquale}, \citenamefont
  {Gianani}, \citenamefont {Sbroscia}, \citenamefont {Booth}, \citenamefont
  {Roccia}, \citenamefont {Raimondi}, \citenamefont {Giovannetti},\ and\
  \citenamefont {Barbieri}}]{Cavina2018}%
  \BibitemOpen
  \bibfield  {author} {\bibinfo {author} {\bibfnamefont {V.}~\bibnamefont
  {Cavina}}, \bibinfo {author} {\bibfnamefont {L.}~\bibnamefont {Mancino}},
  \bibinfo {author} {\bibfnamefont {A.}~\bibnamefont {De~Pasquale}}, \bibinfo
  {author} {\bibfnamefont {I.}~\bibnamefont {Gianani}}, \bibinfo {author}
  {\bibfnamefont {M.}~\bibnamefont {Sbroscia}}, \bibinfo {author}
  {\bibfnamefont {R.~I.}\ \bibnamefont {Booth}}, \bibinfo {author}
  {\bibfnamefont {E.}~\bibnamefont {Roccia}}, \bibinfo {author} {\bibfnamefont
  {R.}~\bibnamefont {Raimondi}}, \bibinfo {author} {\bibfnamefont
  {V.}~\bibnamefont {Giovannetti}},\ and\ \bibinfo {author} {\bibfnamefont
  {M.}~\bibnamefont {Barbieri}},\ }\bibfield  {title} {\bibinfo {title}
  {Bridging thermodynamics and metrology in nonequilibrium quantum
  thermometry},\ }\href {https://doi.org/10.1103/PhysRevA.98.050101} {\bibfield
   {journal} {\bibinfo  {journal} {Phys. Rev. A}\ }\textbf {\bibinfo {volume}
  {98}},\ \bibinfo {pages} {050101(R)} (\bibinfo {year} {2018})}\BibitemShut
  {NoStop}%
\bibitem [{\citenamefont {Seah}\ \emph {et~al.}(2019)\citenamefont {Seah},
  \citenamefont {Nimmrichter}, \citenamefont {Grimmer}, \citenamefont {Santos},
  \citenamefont {Scarani},\ and\ \citenamefont {Landi}}]{Seah2019}%
  \BibitemOpen
  \bibfield  {author} {\bibinfo {author} {\bibfnamefont {S.}~\bibnamefont
  {Seah}}, \bibinfo {author} {\bibfnamefont {S.}~\bibnamefont {Nimmrichter}},
  \bibinfo {author} {\bibfnamefont {D.}~\bibnamefont {Grimmer}}, \bibinfo
  {author} {\bibfnamefont {J.~P.}\ \bibnamefont {Santos}}, \bibinfo {author}
  {\bibfnamefont {V.}~\bibnamefont {Scarani}},\ and\ \bibinfo {author}
  {\bibfnamefont {G.~T.}\ \bibnamefont {Landi}},\ }\bibfield  {title} {\bibinfo
  {title} {Collisional quantum thermometry},\ }\href
  {https://doi.org/10.1103/PhysRevLett.123.180602} {\bibfield  {journal}
  {\bibinfo  {journal} {Phys. Rev. Lett.}\ }\textbf {\bibinfo {volume} {123}},\
  \bibinfo {pages} {180602} (\bibinfo {year} {2019})}\BibitemShut {NoStop}%
\bibitem [{\citenamefont {Toyli}\ \emph {et~al.}(2013)\citenamefont {Toyli},
  \citenamefont {de~las Casas}, \citenamefont {Christle}, \citenamefont
  {Dobrovitski},\ and\ \citenamefont {Awschalom}}]{Toyli8417}%
  \BibitemOpen
  \bibfield  {author} {\bibinfo {author} {\bibfnamefont {D.~M.}\ \bibnamefont
  {Toyli}}, \bibinfo {author} {\bibfnamefont {C.~F.}\ \bibnamefont {de~las
  Casas}}, \bibinfo {author} {\bibfnamefont {D.~J.}\ \bibnamefont {Christle}},
  \bibinfo {author} {\bibfnamefont {V.~V.}\ \bibnamefont {Dobrovitski}},\ and\
  \bibinfo {author} {\bibfnamefont {D.~D.}\ \bibnamefont {Awschalom}},\
  }\bibfield  {title} {\bibinfo {title} {Fluorescence thermometry enhanced by
  the quantum coherence of single spins in diamond},\ }\href
  {https://doi.org/10.1073/pnas.1306825110} {\bibfield  {journal} {\bibinfo
  {journal} {Proc. Natl. Acad. Sci. USA}\ }\textbf {\bibinfo {volume} {110}},\
  \bibinfo {pages} {8417} (\bibinfo {year} {2013})}\BibitemShut {NoStop}%
\bibitem [{\citenamefont {Kucsko}\ \emph {et~al.}(2013)\citenamefont {Kucsko},
  \citenamefont {Maurer}, \citenamefont {Yao}, \citenamefont {Kubo},
  \citenamefont {Noh}, \citenamefont {Lo}, \citenamefont {Park},\ and\
  \citenamefont {Lukin}}]{Kucsko2013}%
  \BibitemOpen
  \bibfield  {author} {\bibinfo {author} {\bibfnamefont {G.}~\bibnamefont
  {Kucsko}}, \bibinfo {author} {\bibfnamefont {P.~C.}\ \bibnamefont {Maurer}},
  \bibinfo {author} {\bibfnamefont {N.~Y.}\ \bibnamefont {Yao}}, \bibinfo
  {author} {\bibfnamefont {M.}~\bibnamefont {Kubo}}, \bibinfo {author}
  {\bibfnamefont {H.~J.}\ \bibnamefont {Noh}}, \bibinfo {author} {\bibfnamefont
  {P.~K.}\ \bibnamefont {Lo}}, \bibinfo {author} {\bibfnamefont
  {H.}~\bibnamefont {Park}},\ and\ \bibinfo {author} {\bibfnamefont {M.~D.}\
  \bibnamefont {Lukin}},\ }\bibfield  {title} {\bibinfo {title}
  {Nanometre-scale thermometry in a living cell},\ }\href
  {https://doi.org/10.1038/nature12373} {\bibfield  {journal} {\bibinfo
  {journal} {Nature}\ }\textbf {\bibinfo {volume} {500}},\ \bibinfo {pages}
  {54} (\bibinfo {year} {2013})}\BibitemShut {NoStop}%
\bibitem [{\citenamefont {Grover}\ \emph {et~al.}(2015)\citenamefont {Grover},
  \citenamefont {Solano}, \citenamefont {Orozco},\ and\ \citenamefont
  {Rolston}}]{PhysRevA.92.013850}%
  \BibitemOpen
  \bibfield  {author} {\bibinfo {author} {\bibfnamefont {J.~A.}\ \bibnamefont
  {Grover}}, \bibinfo {author} {\bibfnamefont {P.}~\bibnamefont {Solano}},
  \bibinfo {author} {\bibfnamefont {L.~A.}\ \bibnamefont {Orozco}},\ and\
  \bibinfo {author} {\bibfnamefont {S.~L.}\ \bibnamefont {Rolston}},\
  }\bibfield  {title} {\bibinfo {title} {Photon-correlation measurements of
  atomic-cloud temperature using an optical nanofiber},\ }\href
  {https://doi.org/10.1103/PhysRevA.92.013850} {\bibfield  {journal} {\bibinfo
  {journal} {Phys. Rev. A}\ }\textbf {\bibinfo {volume} {92}},\ \bibinfo
  {pages} {013850} (\bibinfo {year} {2015})}\BibitemShut {NoStop}%
\bibitem [{\citenamefont {Purdy}\ \emph {et~al.}(2017)\citenamefont {Purdy},
  \citenamefont {Grutter}, \citenamefont {Srinivasan},\ and\ \citenamefont
  {Taylor}}]{purdy2017quantum}%
  \BibitemOpen
  \bibfield  {author} {\bibinfo {author} {\bibfnamefont {T.~P.}\ \bibnamefont
  {Purdy}}, \bibinfo {author} {\bibfnamefont {K.~E.}\ \bibnamefont {Grutter}},
  \bibinfo {author} {\bibfnamefont {K.}~\bibnamefont {Srinivasan}},\ and\
  \bibinfo {author} {\bibfnamefont {J.~M.}\ \bibnamefont {Taylor}},\ }\bibfield
   {title} {\bibinfo {title} {Quantum correlations from a room-temperature
  optomechanical cavity},\ }\href {https://doi.org/10.1126/science.aag1407}
  {\bibfield  {journal} {\bibinfo  {journal} {Science}\ }\textbf {\bibinfo
  {volume} {356}},\ \bibinfo {pages} {1265} (\bibinfo {year}
  {2017})}\BibitemShut {NoStop}%
\bibitem [{\citenamefont {Haupt}\ \emph {et~al.}(2014)\citenamefont {Haupt},
  \citenamefont {Imamoglu},\ and\ \citenamefont
  {Kroner}}]{PhysRevApplied.2.024001}%
  \BibitemOpen
  \bibfield  {author} {\bibinfo {author} {\bibfnamefont {F.}~\bibnamefont
  {Haupt}}, \bibinfo {author} {\bibfnamefont {A.}~\bibnamefont {Imamoglu}},\
  and\ \bibinfo {author} {\bibfnamefont {M.}~\bibnamefont {Kroner}},\
  }\bibfield  {title} {\bibinfo {title} {Single quantum dot as an optical
  thermometer for millikelvin temperatures},\ }\href
  {https://doi.org/10.1103/PhysRevApplied.2.024001} {\bibfield  {journal}
  {\bibinfo  {journal} {Phys. Rev. Applied}\ }\textbf {\bibinfo {volume} {2}},\
  \bibinfo {pages} {024001} (\bibinfo {year} {2014})}\BibitemShut {NoStop}%
\bibitem [{\citenamefont {De~Pasquale}\ \emph {et~al.}(2016)\citenamefont
  {De~Pasquale}, \citenamefont {Rossini}, \citenamefont {Fazio},\ and\
  \citenamefont {Giovannetti}}]{Pasquale2016}%
  \BibitemOpen
  \bibfield  {author} {\bibinfo {author} {\bibfnamefont {A.}~\bibnamefont
  {De~Pasquale}}, \bibinfo {author} {\bibfnamefont {D.}~\bibnamefont
  {Rossini}}, \bibinfo {author} {\bibfnamefont {R.}~\bibnamefont {Fazio}},\
  and\ \bibinfo {author} {\bibfnamefont {V.}~\bibnamefont {Giovannetti}},\
  }\bibfield  {title} {\bibinfo {title} {Local quantum thermal
  susceptibility},\ }\href {https://doi.org/10.1038/ncomms12782} {\bibfield
  {journal} {\bibinfo  {journal} {Nat. Commun.}\ }\textbf {\bibinfo {volume}
  {7}},\ \bibinfo {pages} {12782} (\bibinfo {year} {2016})}\BibitemShut
  {NoStop}%
\bibitem [{\citenamefont {De~Palma}\ \emph {et~al.}(2017)\citenamefont
  {De~Palma}, \citenamefont {De~Pasquale},\ and\ \citenamefont
  {Giovannetti}}]{PhysRevA.95.052115}%
  \BibitemOpen
  \bibfield  {author} {\bibinfo {author} {\bibfnamefont {G.}~\bibnamefont
  {De~Palma}}, \bibinfo {author} {\bibfnamefont {A.}~\bibnamefont
  {De~Pasquale}},\ and\ \bibinfo {author} {\bibfnamefont {V.}~\bibnamefont
  {Giovannetti}},\ }\bibfield  {title} {\bibinfo {title} {Universal locality of
  quantum thermal susceptibility},\ }\href
  {https://doi.org/10.1103/PhysRevA.95.052115} {\bibfield  {journal} {\bibinfo
  {journal} {Phys. Rev. A}\ }\textbf {\bibinfo {volume} {95}},\ \bibinfo
  {pages} {052115} (\bibinfo {year} {2017})}\BibitemShut {NoStop}%
\bibitem [{\citenamefont {Paris}(2015)}]{Paris_2015}%
  \BibitemOpen
  \bibfield  {author} {\bibinfo {author} {\bibfnamefont {M.~G.~A.}\
  \bibnamefont {Paris}},\ }\bibfield  {title} {\bibinfo {title} {Achieving the
  landau bound to precision of quantum thermometry in systems with vanishing
  gap},\ }\href {https://doi.org/10.1088/1751-8113/49/3/03lt02} {\bibfield
  {journal} {\bibinfo  {journal} {J. Phys. A: Math. Theor.}\ }\textbf {\bibinfo
  {volume} {49}},\ \bibinfo {pages} {03LT02} (\bibinfo {year}
  {2015})}\BibitemShut {NoStop}%
\bibitem [{\citenamefont {Degen}\ \emph {et~al.}(2017)\citenamefont {Degen},
  \citenamefont {Reinhard},\ and\ \citenamefont
  {Cappellaro}}]{RevModPhys.89.035002}%
  \BibitemOpen
  \bibfield  {author} {\bibinfo {author} {\bibfnamefont {C.~L.}\ \bibnamefont
  {Degen}}, \bibinfo {author} {\bibfnamefont {F.}~\bibnamefont {Reinhard}},\
  and\ \bibinfo {author} {\bibfnamefont {P.}~\bibnamefont {Cappellaro}},\
  }\bibfield  {title} {\bibinfo {title} {Quantum sensing},\ }\href
  {https://doi.org/10.1103/RevModPhys.89.035002} {\bibfield  {journal}
  {\bibinfo  {journal} {Rev. Mod. Phys.}\ }\textbf {\bibinfo {volume} {89}},\
  \bibinfo {pages} {035002} (\bibinfo {year} {2017})}\BibitemShut {NoStop}%
\bibitem [{\citenamefont {Pezz\`e}\ \emph {et~al.}(2018)\citenamefont
  {Pezz\`e}, \citenamefont {Smerzi}, \citenamefont {Oberthaler}, \citenamefont
  {Schmied},\ and\ \citenamefont {Treutlein}}]{RevModPhys.90.035005}%
  \BibitemOpen
  \bibfield  {author} {\bibinfo {author} {\bibfnamefont {L.}~\bibnamefont
  {Pezz\`e}}, \bibinfo {author} {\bibfnamefont {A.}~\bibnamefont {Smerzi}},
  \bibinfo {author} {\bibfnamefont {M.~K.}\ \bibnamefont {Oberthaler}},
  \bibinfo {author} {\bibfnamefont {R.}~\bibnamefont {Schmied}},\ and\ \bibinfo
  {author} {\bibfnamefont {P.}~\bibnamefont {Treutlein}},\ }\bibfield  {title}
  {\bibinfo {title} {Quantum metrology with nonclassical states of atomic
  ensembles},\ }\href {https://doi.org/10.1103/RevModPhys.90.035005} {\bibfield
   {journal} {\bibinfo  {journal} {Rev. Mod. Phys.}\ }\textbf {\bibinfo
  {volume} {90}},\ \bibinfo {pages} {035005} (\bibinfo {year}
  {2018})}\BibitemShut {NoStop}%
\bibitem [{\citenamefont {Braun}\ \emph {et~al.}(2018)\citenamefont {Braun},
  \citenamefont {Adesso}, \citenamefont {Benatti}, \citenamefont {Floreanini},
  \citenamefont {Marzolino}, \citenamefont {Mitchell},\ and\ \citenamefont
  {Pirandola}}]{RevModPhys.90.035006}%
  \BibitemOpen
  \bibfield  {author} {\bibinfo {author} {\bibfnamefont {D.}~\bibnamefont
  {Braun}}, \bibinfo {author} {\bibfnamefont {G.}~\bibnamefont {Adesso}},
  \bibinfo {author} {\bibfnamefont {F.}~\bibnamefont {Benatti}}, \bibinfo
  {author} {\bibfnamefont {R.}~\bibnamefont {Floreanini}}, \bibinfo {author}
  {\bibfnamefont {U.}~\bibnamefont {Marzolino}}, \bibinfo {author}
  {\bibfnamefont {M.~W.}\ \bibnamefont {Mitchell}},\ and\ \bibinfo {author}
  {\bibfnamefont {S.}~\bibnamefont {Pirandola}},\ }\bibfield  {title} {\bibinfo
  {title} {Quantum-enhanced measurements without entanglement},\ }\href
  {https://doi.org/10.1103/RevModPhys.90.035006} {\bibfield  {journal}
  {\bibinfo  {journal} {Rev. Mod. Phys.}\ }\textbf {\bibinfo {volume} {90}},\
  \bibinfo {pages} {035006} (\bibinfo {year} {2018})}\BibitemShut {NoStop}%
\bibitem [{\citenamefont {Liu}\ \emph {et~al.}(2019)\citenamefont {Liu},
  \citenamefont {Yuan}, \citenamefont {Lu},\ and\ \citenamefont
  {Wang}}]{Liu_2019}%
  \BibitemOpen
  \bibfield  {author} {\bibinfo {author} {\bibfnamefont {J.}~\bibnamefont
  {Liu}}, \bibinfo {author} {\bibfnamefont {H.}~\bibnamefont {Yuan}}, \bibinfo
  {author} {\bibfnamefont {X.-M.}\ \bibnamefont {Lu}},\ and\ \bibinfo {author}
  {\bibfnamefont {X.}~\bibnamefont {Wang}},\ }\bibfield  {title} {\bibinfo
  {title} {Quantum {F}isher information matrix and multiparameter estimation},\
  }\href {https://doi.org/10.1088/1751-8121/ab5d4d} {\bibfield  {journal}
  {\bibinfo  {journal} {J. Phys. A: Math. Theor.}\ }\textbf {\bibinfo {volume}
  {53}},\ \bibinfo {pages} {023001} (\bibinfo {year} {2019})}\BibitemShut
  {NoStop}%
\bibitem [{\citenamefont {Wang}\ \emph {et~al.}(2017)\citenamefont {Wang},
  \citenamefont {Chen},\ and\ \citenamefont {An}}]{Wang_2017}%
  \BibitemOpen
  \bibfield  {author} {\bibinfo {author} {\bibfnamefont {Y.-S.}\ \bibnamefont
  {Wang}}, \bibinfo {author} {\bibfnamefont {C.}~\bibnamefont {Chen}},\ and\
  \bibinfo {author} {\bibfnamefont {J.-H.}\ \bibnamefont {An}},\ }\bibfield
  {title} {\bibinfo {title} {Quantum metrology in local dissipative
  environments},\ }\href {https://doi.org/10.1088/1367-2630/aa8b01} {\bibfield
  {journal} {\bibinfo  {journal} {New J. Phys.}\ }\textbf {\bibinfo {volume}
  {19}},\ \bibinfo {pages} {113019} (\bibinfo {year} {2017})}\BibitemShut
  {NoStop}%
\bibitem [{\citenamefont {Hosten}\ \emph {et~al.}(2016)\citenamefont {Hosten},
  \citenamefont {Engelsen}, \citenamefont {Krishnakumar},\ and\ \citenamefont
  {Kasevich}}]{Hosten}%
  \BibitemOpen
  \bibfield  {author} {\bibinfo {author} {\bibfnamefont {O.}~\bibnamefont
  {Hosten}}, \bibinfo {author} {\bibfnamefont {N.~J.}\ \bibnamefont
  {Engelsen}}, \bibinfo {author} {\bibfnamefont {R.}~\bibnamefont
  {Krishnakumar}},\ and\ \bibinfo {author} {\bibfnamefont {M.~A.}\ \bibnamefont
  {Kasevich}},\ }\bibfield  {title} {\bibinfo {title} {Measurement noise 100
  times lower than the quantum-projection limit using entangled atoms},\ }\href
  {https://doi.org/10.1038/nature16176} {\bibfield  {journal} {\bibinfo
  {journal} {Nature (London)}\ }\textbf {\bibinfo {volume} {529}},\ \bibinfo
  {pages} {505} (\bibinfo {year} {2016})}\BibitemShut {NoStop}%
\bibitem [{\citenamefont {Tse}\ \emph {et~al.}(2019)\citenamefont {Tse} \emph
  {et~al.}}]{PhysRevLett.123.231107}%
  \BibitemOpen
  \bibfield  {author} {\bibinfo {author} {\bibfnamefont {M.}~\bibnamefont
  {Tse}} \emph {et~al.},\ }\bibfield  {title} {\bibinfo {title}
  {Quantum-enhanced advanced ligo detectors in the era of gravitational-wave
  astronomy},\ }\href {https://doi.org/10.1103/PhysRevLett.123.231107}
  {\bibfield  {journal} {\bibinfo  {journal} {Phys. Rev. Lett.}\ }\textbf
  {\bibinfo {volume} {123}},\ \bibinfo {pages} {231107} (\bibinfo {year}
  {2019})}\BibitemShut {NoStop}%
\bibitem [{\citenamefont {Acernese}\ \emph {et~al.}(2019)\citenamefont
  {Acernese} \emph {et~al.}}]{PhysRevLett.123.231108}%
  \BibitemOpen
  \bibfield  {author} {\bibinfo {author} {\bibfnamefont {F.}~\bibnamefont
  {Acernese}} \emph {et~al.} (\bibinfo {collaboration} {Virgo Collaboration}),\
  }\bibfield  {title} {\bibinfo {title} {Increasing the astrophysical reach of
  the advanced virgo detector via the application of squeezed vacuum states of
  light},\ }\href {https://doi.org/10.1103/PhysRevLett.123.231108} {\bibfield
  {journal} {\bibinfo  {journal} {Phys. Rev. Lett.}\ }\textbf {\bibinfo
  {volume} {123}},\ \bibinfo {pages} {231108} (\bibinfo {year}
  {2019})}\BibitemShut {NoStop}%
\bibitem [{\citenamefont {Yu}\ \emph {et~al.}(2020)\citenamefont {Yu},
  \citenamefont {Qin}, \citenamefont {Qin}, \citenamefont {Wang}, \citenamefont
  {Yan}, \citenamefont {Jia},\ and\ \citenamefont
  {Peng}}]{PhysRevApplied.13.024037}%
  \BibitemOpen
  \bibfield  {author} {\bibinfo {author} {\bibfnamefont {J.}~\bibnamefont
  {Yu}}, \bibinfo {author} {\bibfnamefont {Y.}~\bibnamefont {Qin}}, \bibinfo
  {author} {\bibfnamefont {J.}~\bibnamefont {Qin}}, \bibinfo {author}
  {\bibfnamefont {H.}~\bibnamefont {Wang}}, \bibinfo {author} {\bibfnamefont
  {Z.}~\bibnamefont {Yan}}, \bibinfo {author} {\bibfnamefont {X.}~\bibnamefont
  {Jia}},\ and\ \bibinfo {author} {\bibfnamefont {K.}~\bibnamefont {Peng}},\
  }\bibfield  {title} {\bibinfo {title} {Quantum enhanced optical phase
  estimation with a squeezed thermal state},\ }\href
  {https://doi.org/10.1103/PhysRevApplied.13.024037} {\bibfield  {journal}
  {\bibinfo  {journal} {Phys. Rev. Applied}\ }\textbf {\bibinfo {volume}
  {13}},\ \bibinfo {pages} {024037} (\bibinfo {year} {2020})}\BibitemShut
  {NoStop}%
\bibitem [{\citenamefont {Bai}\ \emph {et~al.}(2019)\citenamefont {Bai},
  \citenamefont {Peng}, \citenamefont {Luo},\ and\ \citenamefont
  {An}}]{PhysRevLett.123.040402}%
  \BibitemOpen
  \bibfield  {author} {\bibinfo {author} {\bibfnamefont {K.}~\bibnamefont
  {Bai}}, \bibinfo {author} {\bibfnamefont {Z.}~\bibnamefont {Peng}}, \bibinfo
  {author} {\bibfnamefont {H.-G.}\ \bibnamefont {Luo}},\ and\ \bibinfo {author}
  {\bibfnamefont {J.-H.}\ \bibnamefont {An}},\ }\bibfield  {title} {\bibinfo
  {title} {Retrieving ideal precision in noisy quantum optical metrology},\
  }\href {https://doi.org/10.1103/PhysRevLett.123.040402} {\bibfield  {journal}
  {\bibinfo  {journal} {Phys. Rev. Lett.}\ }\textbf {\bibinfo {volume} {123}},\
  \bibinfo {pages} {040402} (\bibinfo {year} {2019})}\BibitemShut {NoStop}%
\bibitem [{\citenamefont {Fiderer}\ and\ \citenamefont
  {Braun}(2018)}]{Fiderer2018}%
  \BibitemOpen
  \bibfield  {author} {\bibinfo {author} {\bibfnamefont {L.~J.}\ \bibnamefont
  {Fiderer}}\ and\ \bibinfo {author} {\bibfnamefont {D.}~\bibnamefont
  {Braun}},\ }\bibfield  {title} {\bibinfo {title} {Quantum metrology with
  quantum-chaotic sensors},\ }\href
  {https://doi.org/10.1038/s41467-018-03623-z} {\bibfield  {journal} {\bibinfo
  {journal} {Nat. Commun.}\ }\textbf {\bibinfo {volume} {9}},\ \bibinfo {pages}
  {1351} (\bibinfo {year} {2018})}\BibitemShut {NoStop}%
\bibitem [{\citenamefont {Zanardi}\ \emph {et~al.}(2008)\citenamefont
  {Zanardi}, \citenamefont {Paris},\ and\ \citenamefont
  {Campos~Venuti}}]{PhysRevA.78.042105}%
  \BibitemOpen
  \bibfield  {author} {\bibinfo {author} {\bibfnamefont {P.}~\bibnamefont
  {Zanardi}}, \bibinfo {author} {\bibfnamefont {M.~G.~A.}\ \bibnamefont
  {Paris}},\ and\ \bibinfo {author} {\bibfnamefont {L.}~\bibnamefont
  {Campos~Venuti}},\ }\bibfield  {title} {\bibinfo {title} {Quantum criticality
  as a resource for quantum estimation},\ }\href
  {https://doi.org/10.1103/PhysRevA.78.042105} {\bibfield  {journal} {\bibinfo
  {journal} {Phys. Rev. A}\ }\textbf {\bibinfo {volume} {78}},\ \bibinfo
  {pages} {042105} (\bibinfo {year} {2008})}\BibitemShut {NoStop}%
\bibitem [{\citenamefont {Fr\'erot}\ and\ \citenamefont
  {Roscilde}(2018)}]{PhysRevLett.121.020402}%
  \BibitemOpen
  \bibfield  {author} {\bibinfo {author} {\bibfnamefont {I.}~\bibnamefont
  {Fr\'erot}}\ and\ \bibinfo {author} {\bibfnamefont {T.}~\bibnamefont
  {Roscilde}},\ }\bibfield  {title} {\bibinfo {title} {Quantum critical
  metrology},\ }\href {https://doi.org/10.1103/PhysRevLett.121.020402}
  {\bibfield  {journal} {\bibinfo  {journal} {Phys. Rev. Lett.}\ }\textbf
  {\bibinfo {volume} {121}},\ \bibinfo {pages} {020402} (\bibinfo {year}
  {2018})}\BibitemShut {NoStop}%
\bibitem [{\citenamefont {Salado-Mej{\'{\i}}a}\ \emph
  {et~al.}(2021)\citenamefont {Salado-Mej{\'{\i}}a}, \citenamefont
  {Rom{\'{a}}n-Ancheyta}, \citenamefont {Soto-Eguibar},\ and\ \citenamefont
  {Moya-Cessa}}]{Salado_Mej_a_2021}%
  \BibitemOpen
  \bibfield  {author} {\bibinfo {author} {\bibfnamefont {M.}~\bibnamefont
  {Salado-Mej{\'{\i}}a}}, \bibinfo {author} {\bibfnamefont {R.}~\bibnamefont
  {Rom{\'{a}}n-Ancheyta}}, \bibinfo {author} {\bibfnamefont {F.}~\bibnamefont
  {Soto-Eguibar}},\ and\ \bibinfo {author} {\bibfnamefont {H.~M.}\ \bibnamefont
  {Moya-Cessa}},\ }\bibfield  {title} {\bibinfo {title} {Spectroscopy and
  critical quantum thermometry in the ultrastrong coupling regime},\ }\href
  {https://doi.org/10.1088/2058-9565/abdca5} {\bibfield  {journal} {\bibinfo
  {journal} {Quantum Sci. Technol.}\ }\textbf {\bibinfo {volume} {6}},\
  \bibinfo {pages} {025010} (\bibinfo {year} {2021})}\BibitemShut {NoStop}%
\bibitem [{\citenamefont {Aspelmeyer}\ \emph {et~al.}(2014)\citenamefont
  {Aspelmeyer}, \citenamefont {Kippenberg},\ and\ \citenamefont
  {Marquardt}}]{Aspelmeyer2014}%
  \BibitemOpen
  \bibfield  {author} {\bibinfo {author} {\bibfnamefont {M.}~\bibnamefont
  {Aspelmeyer}}, \bibinfo {author} {\bibfnamefont {T.~J.}\ \bibnamefont
  {Kippenberg}},\ and\ \bibinfo {author} {\bibfnamefont {F.}~\bibnamefont
  {Marquardt}},\ }\bibfield  {title} {\bibinfo {title} {Cavity optomechanics},\
  }\href {https://doi.org/10.1103/RevModPhys.86.1391} {\bibfield  {journal}
  {\bibinfo  {journal} {Rev. Mod. Phys.}\ }\textbf {\bibinfo {volume} {86}},\
  \bibinfo {pages} {1391} (\bibinfo {year} {2014})}\BibitemShut {NoStop}%
\bibitem [{\citenamefont {Leggett}\ \emph {et~al.}(1987)\citenamefont
  {Leggett}, \citenamefont {Chakravarty}, \citenamefont {Dorsey}, \citenamefont
  {Fisher}, \citenamefont {Garg},\ and\ \citenamefont
  {Zwerger}}]{RevModPhys.59.1}%
  \BibitemOpen
  \bibfield  {author} {\bibinfo {author} {\bibfnamefont {A.~J.}\ \bibnamefont
  {Leggett}}, \bibinfo {author} {\bibfnamefont {S.}~\bibnamefont
  {Chakravarty}}, \bibinfo {author} {\bibfnamefont {A.~T.}\ \bibnamefont
  {Dorsey}}, \bibinfo {author} {\bibfnamefont {M.~P.~A.}\ \bibnamefont
  {Fisher}}, \bibinfo {author} {\bibfnamefont {A.}~\bibnamefont {Garg}},\ and\
  \bibinfo {author} {\bibfnamefont {W.}~\bibnamefont {Zwerger}},\ }\bibfield
  {title} {\bibinfo {title} {Dynamics of the dissipative two-state system},\
  }\href {https://doi.org/10.1103/RevModPhys.59.1} {\bibfield  {journal}
  {\bibinfo  {journal} {Rev. Mod. Phys.}\ }\textbf {\bibinfo {volume} {59}},\
  \bibinfo {pages} {1} (\bibinfo {year} {1987})}\BibitemShut {NoStop}%
\bibitem [{\citenamefont {Tong}\ and\ \citenamefont
  {Vojta}(2006)}]{PhysRevLett.97.016802}%
  \BibitemOpen
  \bibfield  {author} {\bibinfo {author} {\bibfnamefont {N.-H.}\ \bibnamefont
  {Tong}}\ and\ \bibinfo {author} {\bibfnamefont {M.}~\bibnamefont {Vojta}},\
  }\bibfield  {title} {\bibinfo {title} {Signatures of a noise-induced quantum
  phase transition in a mesoscopic metal ring},\ }\href
  {https://doi.org/10.1103/PhysRevLett.97.016802} {\bibfield  {journal}
  {\bibinfo  {journal} {Phys. Rev. Lett.}\ }\textbf {\bibinfo {volume} {97}},\
  \bibinfo {pages} {016802} (\bibinfo {year} {2006})}\BibitemShut {NoStop}%
\bibitem [{\citenamefont {Forn-D{\'i}az}\ \emph {et~al.}(2017)\citenamefont
  {Forn-D{\'i}az}, \citenamefont {Garc{\'i}a-Ripoll}, \citenamefont
  {Peropadre}, \citenamefont {Orgiazzi}, \citenamefont {Yurtalan},
  \citenamefont {Belyansky}, \citenamefont {Wilson},\ and\ \citenamefont
  {Lupascu}}]{Forn-Diaz2017}%
  \BibitemOpen
  \bibfield  {author} {\bibinfo {author} {\bibfnamefont {P.}~\bibnamefont
  {Forn-D{\'i}az}}, \bibinfo {author} {\bibfnamefont {J.~J.}\ \bibnamefont
  {Garc{\'i}a-Ripoll}}, \bibinfo {author} {\bibfnamefont {B.}~\bibnamefont
  {Peropadre}}, \bibinfo {author} {\bibfnamefont {J.-L.}\ \bibnamefont
  {Orgiazzi}}, \bibinfo {author} {\bibfnamefont {M.~A.}\ \bibnamefont
  {Yurtalan}}, \bibinfo {author} {\bibfnamefont {R.}~\bibnamefont {Belyansky}},
  \bibinfo {author} {\bibfnamefont {C.~M.}\ \bibnamefont {Wilson}},\ and\
  \bibinfo {author} {\bibfnamefont {A.}~\bibnamefont {Lupascu}},\ }\bibfield
  {title} {\bibinfo {title} {Ultrastrong coupling of a single artificial atom
  to an electromagnetic continuum in the nonperturbative regime},\ }\href
  {https://doi.org/10.1038/nphys3905} {\bibfield  {journal} {\bibinfo
  {journal} {Nature Physics}\ }\textbf {\bibinfo {volume} {13}},\ \bibinfo
  {pages} {39} (\bibinfo {year} {2017})}\BibitemShut {NoStop}%
\bibitem [{\citenamefont {Paladino}\ \emph {et~al.}(2014)\citenamefont
  {Paladino}, \citenamefont {Galperin}, \citenamefont {Falci},\ and\
  \citenamefont {Altshuler}}]{RevModPhys.86.361}%
  \BibitemOpen
  \bibfield  {author} {\bibinfo {author} {\bibfnamefont {E.}~\bibnamefont
  {Paladino}}, \bibinfo {author} {\bibfnamefont {Y.~M.}\ \bibnamefont
  {Galperin}}, \bibinfo {author} {\bibfnamefont {G.}~\bibnamefont {Falci}},\
  and\ \bibinfo {author} {\bibfnamefont {B.~L.}\ \bibnamefont {Altshuler}},\
  }\bibfield  {title} {\bibinfo {title} {$1/f$ noise: Implications for
  solid-state quantum information},\ }\href
  {https://doi.org/10.1103/RevModPhys.86.361} {\bibfield  {journal} {\bibinfo
  {journal} {Rev. Mod. Phys.}\ }\textbf {\bibinfo {volume} {86}},\ \bibinfo
  {pages} {361} (\bibinfo {year} {2014})}\BibitemShut {NoStop}%
\bibitem [{\citenamefont {Porras}\ \emph {et~al.}(2008)\citenamefont {Porras},
  \citenamefont {Marquardt}, \citenamefont {von Delft},\ and\ \citenamefont
  {Cirac}}]{PhysRevA.78.010101}%
  \BibitemOpen
  \bibfield  {author} {\bibinfo {author} {\bibfnamefont {D.}~\bibnamefont
  {Porras}}, \bibinfo {author} {\bibfnamefont {F.}~\bibnamefont {Marquardt}},
  \bibinfo {author} {\bibfnamefont {J.}~\bibnamefont {von Delft}},\ and\
  \bibinfo {author} {\bibfnamefont {J.~I.}\ \bibnamefont {Cirac}},\ }\bibfield
  {title} {\bibinfo {title} {Mesoscopic spin-boson models of trapped ions},\
  }\href {https://doi.org/10.1103/PhysRevA.78.010101} {\bibfield  {journal}
  {\bibinfo  {journal} {Phys. Rev. A}\ }\textbf {\bibinfo {volume} {78}},\
  \bibinfo {pages} {010101(R)} (\bibinfo {year} {2008})}\BibitemShut {NoStop}%
\bibitem [{\citenamefont {Shi}\ \emph {et~al.}(2018)\citenamefont {Shi},
  \citenamefont {Chang},\ and\ \citenamefont
  {Garc\'{\i}a-Ripoll}}]{PhysRevLett.120.153602}%
  \BibitemOpen
  \bibfield  {author} {\bibinfo {author} {\bibfnamefont {T.}~\bibnamefont
  {Shi}}, \bibinfo {author} {\bibfnamefont {Y.}~\bibnamefont {Chang}},\ and\
  \bibinfo {author} {\bibfnamefont {J.~J.}\ \bibnamefont
  {Garc\'{\i}a-Ripoll}},\ }\bibfield  {title} {\bibinfo {title} {Ultrastrong
  coupling few-photon scattering theory},\ }\href
  {https://doi.org/10.1103/PhysRevLett.120.153602} {\bibfield  {journal}
  {\bibinfo  {journal} {Phys. Rev. Lett.}\ }\textbf {\bibinfo {volume} {120}},\
  \bibinfo {pages} {153602} (\bibinfo {year} {2018})}\BibitemShut {NoStop}%
\bibitem [{\citenamefont {Weiss}(2012)}]{Weiss}%
  \BibitemOpen
  \bibfield  {author} {\bibinfo {author} {\bibfnamefont {U.}~\bibnamefont
  {Weiss}},\ }\href@noop {} {\emph {\bibinfo {title} {Quantum dissipative
  systems}}}\ (\bibinfo  {publisher} {World scientific, Singapore},\ \bibinfo
  {year} {2012})\BibitemShut {NoStop}%
\bibitem [{\citenamefont {Feynman}\ and\ \citenamefont
  {Vernon}(1963)}]{Feynman1963}%
  \BibitemOpen
  \bibfield  {author} {\bibinfo {author} {\bibfnamefont {R.~P.}\ \bibnamefont
  {Feynman}}\ and\ \bibinfo {author} {\bibfnamefont {F.~L.}\ \bibnamefont
  {Vernon}},\ }\bibfield  {title} {\bibinfo {title} {The theory of a general
  quantum system interacting with a linear dissipative system},\ }\href
  {https://doi.org/https://doi.org/10.1016/0003-4916(63)90068-X} {\bibfield
  {journal} {\bibinfo  {journal} {Ann. Phys.}\ }\textbf {\bibinfo {volume}
  {24}},\ \bibinfo {pages} {118} (\bibinfo {year} {1963})}\BibitemShut
  {NoStop}%
\bibitem [{\citenamefont {An}\ and\ \citenamefont {Zhang}(2007)}]{An2007}%
  \BibitemOpen
  \bibfield  {author} {\bibinfo {author} {\bibfnamefont {J.-H.}\ \bibnamefont
  {An}}\ and\ \bibinfo {author} {\bibfnamefont {W.-M.}\ \bibnamefont {Zhang}},\
  }\bibfield  {title} {\bibinfo {title} {Non-markovian entanglement dynamics of
  noisy continuous-variable quantum channels},\ }\href
  {https://doi.org/10.1103/PhysRevA.76.042127} {\bibfield  {journal} {\bibinfo
  {journal} {Phys. Rev. A}\ }\textbf {\bibinfo {volume} {76}},\ \bibinfo
  {pages} {042127} (\bibinfo {year} {2007})}\BibitemShut {NoStop}%
\bibitem [{\citenamefont {Yang}\ \emph {et~al.}(2014)\citenamefont {Yang},
  \citenamefont {An}, \citenamefont {Luo}, \citenamefont {Li},\ and\
  \citenamefont {Oh}}]{Yang2014}%
  \BibitemOpen
  \bibfield  {author} {\bibinfo {author} {\bibfnamefont {C.-J.}\ \bibnamefont
  {Yang}}, \bibinfo {author} {\bibfnamefont {J.-H.}\ \bibnamefont {An}},
  \bibinfo {author} {\bibfnamefont {H.-G.}\ \bibnamefont {Luo}}, \bibinfo
  {author} {\bibfnamefont {Y.}~\bibnamefont {Li}},\ and\ \bibinfo {author}
  {\bibfnamefont {C.~H.}\ \bibnamefont {Oh}},\ }\bibfield  {title} {\bibinfo
  {title} {Canonical versus noncanonical equilibration dynamics of open quantum
  systems},\ }\href {https://doi.org/10.1103/PhysRevE.90.022122} {\bibfield
  {journal} {\bibinfo  {journal} {Phys. Rev. E}\ }\textbf {\bibinfo {volume}
  {90}},\ \bibinfo {pages} {022122} (\bibinfo {year} {2014})}\BibitemShut
  {NoStop}%
\bibitem [{\citenamefont {Breuer}\ and\ \citenamefont
  {Petruccione}(2002)}]{breuer2002theory}%
  \BibitemOpen
  \bibfield  {author} {\bibinfo {author} {\bibfnamefont {H.}~\bibnamefont
  {Breuer}}\ and\ \bibinfo {author} {\bibfnamefont {F.}~\bibnamefont
  {Petruccione}},\ }\href@noop {} {\emph {\bibinfo {title} {The Theory of Open
  Quantum Systems}}}\ (\bibinfo  {publisher} {Oxford University Press, New
  York},\ \bibinfo {year} {2002})\BibitemShut {NoStop}%
\bibitem [{\citenamefont {Huelga}\ \emph {et~al.}(1997)\citenamefont {Huelga},
  \citenamefont {Macchiavello}, \citenamefont {Pellizzari}, \citenamefont
  {Ekert}, \citenamefont {Plenio},\ and\ \citenamefont
  {Cirac}}]{PhysRevLett.79.3865}%
  \BibitemOpen
  \bibfield  {author} {\bibinfo {author} {\bibfnamefont {S.~F.}\ \bibnamefont
  {Huelga}}, \bibinfo {author} {\bibfnamefont {C.}~\bibnamefont
  {Macchiavello}}, \bibinfo {author} {\bibfnamefont {T.}~\bibnamefont
  {Pellizzari}}, \bibinfo {author} {\bibfnamefont {A.~K.}\ \bibnamefont
  {Ekert}}, \bibinfo {author} {\bibfnamefont {M.~B.}\ \bibnamefont {Plenio}},\
  and\ \bibinfo {author} {\bibfnamefont {J.~I.}\ \bibnamefont {Cirac}},\
  }\bibfield  {title} {\bibinfo {title} {Improvement of frequency standards
  with quantum entanglement},\ }\href
  {https://doi.org/10.1103/PhysRevLett.79.3865} {\bibfield  {journal} {\bibinfo
   {journal} {Phys. Rev. Lett.}\ }\textbf {\bibinfo {volume} {79}},\ \bibinfo
  {pages} {3865} (\bibinfo {year} {1997})}\BibitemShut {NoStop}%
\bibitem [{\citenamefont {Caves}(1981)}]{PhysRevD.23.1693}%
  \BibitemOpen
  \bibfield  {author} {\bibinfo {author} {\bibfnamefont {C.~M.}\ \bibnamefont
  {Caves}},\ }\bibfield  {title} {\bibinfo {title} {Quantum-mechanical noise in
  an interferometer},\ }\href {https://doi.org/10.1103/PhysRevD.23.1693}
  {\bibfield  {journal} {\bibinfo  {journal} {Phys. Rev. D}\ }\textbf {\bibinfo
  {volume} {23}},\ \bibinfo {pages} {1693} (\bibinfo {year}
  {1981})}\BibitemShut {NoStop}%
\bibitem [{\citenamefont {Wu}\ \emph {et~al.}(2021{\natexlab{a}})\citenamefont
  {Wu}, \citenamefont {Bai},\ and\ \citenamefont {An}}]{Wu2021}%
  \BibitemOpen
  \bibfield  {author} {\bibinfo {author} {\bibfnamefont {W.}~\bibnamefont
  {Wu}}, \bibinfo {author} {\bibfnamefont {S.-Y.}\ \bibnamefont {Bai}},\ and\
  \bibinfo {author} {\bibfnamefont {J.-H.}\ \bibnamefont {An}},\ }\bibfield
  {title} {\bibinfo {title} {Non-markovian sensing of a quantum reservoir},\
  }\href {https://doi.org/10.1103/PhysRevA.103.L010601} {\bibfield  {journal}
  {\bibinfo  {journal} {Phys. Rev. A}\ }\textbf {\bibinfo {volume} {103}},\
  \bibinfo {pages} {L010601} (\bibinfo {year}
  {2021}{\natexlab{a}})}\BibitemShut {NoStop}%
\bibitem [{\citenamefont {Wu}\ \emph {et~al.}(2021{\natexlab{b}})\citenamefont
  {Wu}, \citenamefont {Peng}, \citenamefont {Bai},\ and\ \citenamefont
  {An}}]{PhysRevApplied.15.054042}%
  \BibitemOpen
  \bibfield  {author} {\bibinfo {author} {\bibfnamefont {W.}~\bibnamefont
  {Wu}}, \bibinfo {author} {\bibfnamefont {Z.}~\bibnamefont {Peng}}, \bibinfo
  {author} {\bibfnamefont {S.-Y.}\ \bibnamefont {Bai}},\ and\ \bibinfo {author}
  {\bibfnamefont {J.-H.}\ \bibnamefont {An}},\ }\bibfield  {title} {\bibinfo
  {title} {Threshold for a discrete-variable sensor of quantum reservoirs},\
  }\href {https://doi.org/10.1103/PhysRevApplied.15.054042} {\bibfield
  {journal} {\bibinfo  {journal} {Phys. Rev. Applied}\ }\textbf {\bibinfo
  {volume} {15}},\ \bibinfo {pages} {054042} (\bibinfo {year}
  {2021}{\natexlab{b}})}\BibitemShut {NoStop}%
\bibitem [{\citenamefont {Braunstein}\ and\ \citenamefont {van
  Loock}(2005)}]{RevModPhys.77.513}%
  \BibitemOpen
  \bibfield  {author} {\bibinfo {author} {\bibfnamefont {S.~L.}\ \bibnamefont
  {Braunstein}}\ and\ \bibinfo {author} {\bibfnamefont {P.}~\bibnamefont {van
  Loock}},\ }\bibfield  {title} {\bibinfo {title} {Quantum information with
  continuous variables},\ }\href {https://doi.org/10.1103/RevModPhys.77.513}
  {\bibfield  {journal} {\bibinfo  {journal} {Rev. Mod. Phys.}\ }\textbf
  {\bibinfo {volume} {77}},\ \bibinfo {pages} {513} (\bibinfo {year}
  {2005})}\BibitemShut {NoStop}%
\bibitem [{\citenamefont {{\v{S}}afr{\'{a}}nek}\ \emph
  {et~al.}(2015)\citenamefont {{\v{S}}afr{\'{a}}nek}, \citenamefont {Lee},\
  and\ \citenamefont {Fuentes}}]{_afr_nek_2015}%
  \BibitemOpen
  \bibfield  {author} {\bibinfo {author} {\bibfnamefont {D.}~\bibnamefont
  {{\v{S}}afr{\'{a}}nek}}, \bibinfo {author} {\bibfnamefont {A.~R.}\
  \bibnamefont {Lee}},\ and\ \bibinfo {author} {\bibfnamefont {I.}~\bibnamefont
  {Fuentes}},\ }\bibfield  {title} {\bibinfo {title} {Quantum parameter
  estimation using multi-mode gaussian states},\ }\href
  {https://doi.org/10.1088/1367-2630/17/7/073016} {\bibfield  {journal}
  {\bibinfo  {journal} {New J. Phys.}\ }\textbf {\bibinfo {volume} {17}},\
  \bibinfo {pages} {073016} (\bibinfo {year} {2015})}\BibitemShut {NoStop}%
\bibitem [{\citenamefont {Hovhannisyan}\ and\ \citenamefont
  {Correa}(2018)}]{PhysRevB.98.045101}%
  \BibitemOpen
  \bibfield  {author} {\bibinfo {author} {\bibfnamefont {K.~V.}\ \bibnamefont
  {Hovhannisyan}}\ and\ \bibinfo {author} {\bibfnamefont {L.~A.}\ \bibnamefont
  {Correa}},\ }\bibfield  {title} {\bibinfo {title} {Measuring the temperature
  of cold many-body quantum systems},\ }\href
  {https://doi.org/10.1103/PhysRevB.98.045101} {\bibfield  {journal} {\bibinfo
  {journal} {Phys. Rev. B}\ }\textbf {\bibinfo {volume} {98}},\ \bibinfo
  {pages} {045101} (\bibinfo {year} {2018})}\BibitemShut {NoStop}%
\bibitem [{\citenamefont {Liu}\ and\ \citenamefont {Houck}(2017)}]{Liu2016}%
  \BibitemOpen
  \bibfield  {author} {\bibinfo {author} {\bibfnamefont {Y.}~\bibnamefont
  {Liu}}\ and\ \bibinfo {author} {\bibfnamefont {A.~A.}\ \bibnamefont
  {Houck}},\ }\bibfield  {title} {\bibinfo {title} {Quantum electrodynamics
  near a photonic bandgap},\ }\href {http://dx.doi.org/10.1038/nphys3834}
  {\bibfield  {journal} {\bibinfo  {journal} {Nat. Phys.}\ }\textbf {\bibinfo
  {volume} {13}},\ \bibinfo {pages} {48} (\bibinfo {year} {2017})}\BibitemShut
  {NoStop}%
\bibitem [{\citenamefont {Krinner}\ \emph {et~al.}(2018)\citenamefont
  {Krinner}, \citenamefont {Stewart}, \citenamefont {Pazmi\~{n}o},
  \citenamefont {Kwon},\ and\ \citenamefont {Schneble}}]{Kri2018}%
  \BibitemOpen
  \bibfield  {author} {\bibinfo {author} {\bibfnamefont {L.}~\bibnamefont
  {Krinner}}, \bibinfo {author} {\bibfnamefont {M.}~\bibnamefont {Stewart}},
  \bibinfo {author} {\bibfnamefont {A.}~\bibnamefont {Pazmi\~{n}o}}, \bibinfo
  {author} {\bibfnamefont {J.}~\bibnamefont {Kwon}},\ and\ \bibinfo {author}
  {\bibfnamefont {D.}~\bibnamefont {Schneble}},\ }\bibfield  {title} {\bibinfo
  {title} {Spontaneous emission of matter waves from a tunable open quantum
  system},\ }\href {https://doi.org/10.1038/s41586-018-0348-z} {\bibfield
  {journal} {\bibinfo  {journal} {Nature (London)}\ }\textbf {\bibinfo {volume}
  {559}},\ \bibinfo {pages} {589} (\bibinfo {year} {2018})}\BibitemShut
  {NoStop}%
\bibitem [{\citenamefont {Hassani}(2013)}]{book:1123190}%
  \BibitemOpen
  \bibfield  {author} {\bibinfo {author} {\bibfnamefont {S.}~\bibnamefont
  {Hassani}},\ }\href@noop {} {\emph {\bibinfo {title} {Mathematical Physics: A
  Modern Introduction to Its Foundations}}},\ \bibinfo {edition} {2nd}\ ed.\
  (\bibinfo  {publisher} {Springer International Publishing, New York},\
  \bibinfo {year} {2013})\BibitemShut {NoStop}%
\end{thebibliography}%

\end{document}